\documentclass[12pt,english]{achemso}
\pdfoutput = 1
\usepackage[LGR,T1]{fontenc}
\usepackage{textcomp}
\usepackage[latin9]{inputenc}
\usepackage{color}
\usepackage{array}
\usepackage{booktabs}
\usepackage{multirow}
\usepackage{amsmath}
\usepackage{amssymb}
\usepackage{stackrel}
\usepackage{graphicx}

\makeatletter

\title{\textcolor{black}{Strain-Engineered s-C\textsubscript{3}N\textsubscript{6}
Monolayer for Efficient Water Splitting: A first-principles study}}

\author{\textcolor{black}{Khushboo Dange}}

\email{khushboodange@gmail.com}

\affiliation{\textcolor{black}{Department of Physics, Indian Institute of Technology
Bombay, Powai, Mumbai 400076, India}}

\author{\textcolor{black}{Alok Shukla}}

\email{shukla@iitb.ac.in}

\affiliation{\textcolor{black}{Department of Physics, Indian Institute of Technology
Bombay, Powai, Mumbai 400076, India}}

\DeclareRobustCommand{\greektext}{%
  \fontencoding{LGR}\selectfont\def\encodingdefault{LGR}}
\DeclareRobustCommand{\textgreek}[1]{\leavevmode{\greektext #1}}

\newcommand{\lyxmathsym}[1]{\ifmmode\begingroup\def\b@ld{bold}
  \text{\ifx\math@version\b@ld\bfseries\fi#1}\endgroup\else#1\fi}

\providecommand{\tabularnewline}{\\}

\usepackage{babel}

\makeatother

\usepackage{babel}
\begin{document}
\begin{abstract}
\textcolor{black}{Photocatalytic water splitting offers a sustainable
route for solar-to-hydrogen energy conversion. However, identifying
metal-free semiconductors that simultaneously satisfy stability, electronic,
optical, and band-alignment requirements remains challenging. Here,
we investigate the structural, mechanical, electronic, optical, and
photocatalytic properties of the recently proposed two-dimensional
s-C\textsubscript{3}N\textsubscript{6} monolayer using first-principles
density functional theory. Ab initio molecular dynamics simulations
confirm its thermal stability at ambient and elevated temperatures,
while elastic constant analysis demonstrates mechanical robustness.
Hybrid HSE06 calculations reveal that pristine s-C\textsubscript{3}N\textsubscript{6}
monolayer is a direct-band-gap semiconductor with a gap of 2.62 eV.
However, its conduction-band minimum lies below the hydrogen reduction
potential, preventing spontaneous hydrogen evolution. To overcome
this limitation, we employ strain engineering to modulate its electronic
structure and band-edge alignment. Both biaxial and uniaxial strains
in the range of \textminus 10\% to +10\% are systematically explored.
We find that compressive biaxial strains of \textminus 8\% and \textminus 10\%
uniquely tune the band edges to simultaneously satisfy the thermodynamic
requirements for hydrogen and oxygen evolution reactions, enabling
spontaneous overall water splitting. These photocatalytically active
strain (-8\% and -10\% biaxial strains) states are further shown to
be mechanically and thermally stable, underlining their experimental
feasibility. The calculations of the optical transitions both in the
pristine, and photocatalytically active strained structures, show
that the fundamental electronic gap is optically dark due to symmetry-forbidden
transitions, with the first absorption peak in the UV region, arising
from deeper valence states. Crucially, a strain-induced mobility mismatch
between electrons and holes, arising from divergent effective mass
trends, facilitates efficient charge separation and carrier protection.
However, thermodynamic modeling of surface kinetics reveals that the
s-C\textsubscript{3}N\textsubscript{6} surface binds intermediates
too strongly, necessitating a co-catalyst to overcome the resulting
kinetic barriers. Our results establish strain engineering as an effective
strategy to simultaneously tailor band-edge alignment, carrier dynamics,
and optical transitions in the s-C\textsubscript{3}N\textsubscript{6}
monolayer, highlighting its potential for stable charge-generation
for 2D photocatalytic water splitting.}
\end{abstract}
\textbf{\textcolor{black}{Keywords:}}\textcolor{black}{{} s-C\textsubscript{3\textsuperscript{}}N\textsubscript{6\textsuperscript{}}
monolayer; density functional theory; electronic properties; elastic
properties; optical properties; strain engineering; photocatalysis}

\section{\textcolor{black}{Introduction}}

\textcolor{black}{The escalating global energy demand, coupled with
the severe environmental consequences of carbon emissions, has accelerated
the transition towards clean and sustainable alternatives. The direct
conversion of solar energy into chemical fuels is a promising route
for the sustainable generation of renewable energy, with photocatalysis
emerging as a highly efficient and versatile method to drive this
process. The primary focus in this domain is the generation of hydrogen,
a clean energy carrier with high energy density. Photocatalytic water
splitting is a key reaction that enables hydrogen production from
water, yielding oxygen as the sole byproduct. Achieving efficient
solar-driven water splitting requires a catalyst that can efficiently
absorb sunlight, possesses suitable band alignment to drive both the
hydrogen and oxygen evolution reactions (HER and OER), and enables
fast charge separation while suppressing the recombination of photo-generated
carriers. Consequently, research aimed at finding such an effective
photocatalyst is increasing at a rapid pace.}

\textcolor{black}{The extensive search for effective photocatalysts
has historically focused on various materials, beginning with metal
oxides such as TiO\textsubscript{2} \cite{Chen2007}, metal sulfides
\cite{Zhang2013}, and many other compounds \cite{Kato1999,DOMEN1986,Karthik2019}.
However, two-dimensional (2D) semiconductors are emerging as superior
photocatalysts when compared to their bulk counterparts. This superiority
stems primarily from their exceptional characteristics: a high surface-to-volume
ratio which maximizes the effective area for photon absorption, and
surface-water interaction \cite{LI2019}. Crucially, the reduced thickness
of 2D materials ensures a significantly shorter carrier migration
distance, thereby minimizing their recombination rate and promoting
efficient charge separation \cite{Singh2015}. Furthermore, unlike
their bulk analogs, the electronic structure of 2D materials can be
readily tuned through methods like doping, strain engineering, external
electric fields, or the formation of heterostructures \cite{Dange2023,Li2013,Dange2025,Dinghao2021}.
2D heterostructures play a significant role in advancing photocatalytic
design. For example, Z-scheme systems like PdSSe/GeC and Janus-based
TiBrTe/ZrSeS have been identified through machine learning-guided
screening to exhibit accelerated interfacial charge transfer \cite{Zhang2025,Xi2026}.
Furthermore, advanced electronic and coordination modulation strategies
have been widely explored. These include the aggregation-rebalance
mechanism in porous NiS/PCNs \cite{ZHANGKong2025}, dual-doped P-Ni\textsubscript{1/2}Fe\textsubscript{1/2}
LDH \cite{ZHANG2026}, and charge buffering in dual-atom catalysts
on B-N-doped graphene \cite{HE2026}, which collectively demonstrate
a universal framework for lowering kinetic overpotentials and optimizing
charge transport across diverse catalytic applications.}

\textcolor{black}{Among the 2D \textgreek{\textpi}-conjugated materials,
graphitic carbon nitrides (g-C\textsubscript{x}N\textsubscript{y})
have emerged as highly attractive candidates for photocatalytic applications
due to several synergistic properties. Their inherent \textgreek{\textpi}-conjugation
features delocalized electrons, which not only extend the absorption
edge for wider solar spectrum coverage, but also enable rapid charge
separation \cite{Ong2016}. Various g-C\textsubscript{x}N\textsubscript{y}
monolayers such as g-C\textsubscript{3}N\textsubscript{4} (both
s-triazine and tri-s-triazine) \cite{Algara2014,Chuang2016}, C\textsubscript{2}N
\cite{Mahmood2015}, C\textsubscript{3}N \cite{Javeed2016}, C\textsubscript{3}N\textsubscript{5}
\cite{Kumar2019}, and C\textsubscript{6}N\textsubscript{7} \cite{ZHAO2021}
have been successfully fabricated and demonstrate marked improvements
in photovoltaic and photocatalytic applications \cite{Makaremi2018}.
The s-C\textsubscript{3}N\textsubscript{6 } motif is another structure
within the g-C\textsubscript{x}N\textsubscript{y} family, consisting
of s-triazine units linked to each other via N-N azo-linkages \cite{MORTAZAVI2020}.
Its stable structure was predicted by Mortazavi }\textit{\textcolor{black}{et
al.}}\textcolor{black}{{} \cite{MORTAZAVI2020} based on the structural
framework of C\textsubscript{3}N\textsubscript{5}, which contains
tri-s-triazine units \cite{Kumar2019}. Due to the promising structure
of s-C\textsubscript{3}N\textsubscript{6} monolayer, few studies
have investigated its electronic and optical properties using first-principles
techniques \cite{MORTAZAVI2020,Deshpande2022,SHE2024}. These initial
studies have explored diverse applications; for instance, Deshpande
}\textit{\textcolor{black}{et al.}}\textcolor{black}{{} \cite{Deshpande2022}
demonstrated that the transition-metal embedded s-C\textsubscript{3}N\textsubscript{6}
monolayer possesses selective gas capturing properties and thus has
potential for sensing applications. While the intrinsic electronic
band gap of the s-C\textsubscript{3}N\textsubscript{6} monolayer
(2.62 eV using the HSE06 functional) falls within the visible energy
region, making it seemingly suitable for photocatalytic applications,
pristine s-C\textsubscript{3}N\textsubscript{6} monolayer ultimately
fails to meet the important criteria of appropriate band alignment
with the redox potential levels of water. This deficiency restricts
its application as a photocatalyst for water splitting. Consequently,
She }\textit{\textcolor{black}{et al.}}\textcolor{black}{{} \cite{SHE2024}
demonstrated that s-C\textsubscript{3}N\textsubscript{6} monolayer
satisfies the necessary band alignment criteria for photocatalysis
when doped with Cl atoms.}

\textcolor{black}{While chemical modification (e.g., Cl-doping) resolves
the band alignment issue, this method poses challenges in synthesis,
impurity control, and stability. Therefore, strain engineering, which
is a cleaner and mechanically tailored approach, is a highly advantageous
alternative for optimizing s-C\textsubscript{3}N\textsubscript{6}
monolayer for water splitting. Furthermore, this method is practically
feasible, as layers can get strained when exfoliated onto a substrate
\cite{Metzger2010}. Therefore, in this work, we investigate the influence
of uniaxial and biaxial strains ranging from -10\% to 10\% on the
electronic, optical, and photocatalytic properties of the s-C\textsubscript{3}N\textsubscript{6}
monolayer. Our primary objective is to tune the band alignment for
spontaneous water splitting followed by study of surface kinetics.}

\textcolor{black}{The remainder of this paper is organized as follows.
The next section briefly describes the employed computational methods.
This is followed by a detailed presentation and discussion of our
findings, culminating in the final summary and conclusions.}

\section{\textcolor{black}{Computational Methodology}}

\textcolor{black}{All calculations were performed within the framework
of first-principles density functional theory (DFT) \cite{hohenberg1964,kohn1965}
using the Vienna ab initio simulation package (VASP) \cite{kresse1996,Georg1996}.
The Kohn--Sham wave functions were expanded in a plane-wave basis
with a kinetic energy cutoff of 600 eV. Electron--ion interactions
were treated using the projector augmented-wave (PAW) method \cite{kresse1999,blochl1994},
with valence configurations of $2s^{2}2p^{2}$ for C and $2s^{2}2p^{3}$
for N atoms. A vacuum spacing of 15 $\text{\AA}$ was introduced along
the out-of-plane direction to eliminate interactions between periodic
images of the s-C\textsubscript{3}N\textsubscript{6} monolayer.
The self-consistent field (SCF) calculations to solve the Kohn-Sham
equations were converged to an energy tolerance of $10^{-6}$ eV,
and all atomic positions were relaxed until the Hellmann--Feynman
forces were less than $10^{-2}$ eV/{\AA}. Brillouin zone sampling
was carried out using a }\textbf{\textcolor{black}{$\Gamma$}}\textcolor{black}{-centered}\textbf{\textcolor{black}{{}
}}\textcolor{black}{$15\times15\times1$ }\textbf{\textcolor{black}{k}}\textcolor{black}{-mesh.
Initial structural optimization and preliminary electronic properties
were obtained using the Perdew--Burke--Ernzerhof (PBE) exchange--correlation
functional within the generalized gradient approximation (GGA) \cite{perdew1996}.
To mitigate the well-known underestimation of band gaps by GGA-PBE,
the final electronic band structures were evaluated using the hybrid
Heyd--Scuseria--Ernzerhof (HSE06) functional \cite{Heyd2003}.}

\subsection*{\textcolor{black}{Thermal and elastic stability calculations:}}

\textcolor{black}{The thermal stability was examined via ab initio
molecular dynamics (AIMD) simulations \cite{Kresse1994} within the
canonical (NVT) ensemble using a Nosé-Hoover thermostat at temperatures
of 300 K and 400 K. Elastic properties were evaluated at the HSE06
level by applying small in-plane strains ranging from -2\% to 2\%
in increments of 0.5\%. All deformations were restricted to the harmonic
elastic regime, where the strain-dependent elastic energy $\Delta E(V,\{\epsilon_{i}\})$
is given by}

\textcolor{black}{
\begin{equation}
\Delta E(V,\{\epsilon_{i}\})=E(V,\{\epsilon_{i}\})-E(V_{0},0)=\frac{V_{0}}{2}\sum_{i,j=1}^{6}C_{ij}\epsilon_{i}\epsilon_{j},\label{eq:energy-strain}
\end{equation}
}

\textcolor{black}{where $E(V,\{\epsilon_{i}\})$ and $E(V_{0},0)$
represent the total energies of the strained and equilibrium configurations
with volumes $V$ and $V_{0}$, respectively, and $C_{ij}$ denotes
the elastic stiffness constants. For a 2D system, the $6\times6$
elastic stiffness tensor reduces to a $3\times3$ symmetric matrix
involving in-plane components ($i,j$= 1, 2, 6). These elastic constants
were extracted using the energy--strain relations as implemented
in the VASPKIT package \cite{WANG2021}. Owing to the hexagonal symmetry
of the $s-C_{3}N_{6}$ monolayer, only two independent elastic constants
exist ($C_{11}$ and $C_{12}$), satisfying the relations $C_{11}=C_{22}$,
$C_{12}=C_{21}$, and $C_{66}=\frac{1}{2}(C_{11}-C_{12})$. Additionally,
the in-plane orientation-dependent Young\textquoteright s modulus
$E(\theta)$ and the Poisson's ratio $\nu(\theta)$ were calculated
using the following standard expressions derived from the elastic
tensor
\begin{equation}
E(\theta)=\frac{C_{11}C_{22}-C_{12}^{2}}{C_{11}\sin^{4}\theta+\left(\frac{C_{11}C_{22}-C_{12}^{2}}{C_{66}}-2C_{12}\right)\sin^{2}\theta\cos^{2}\theta+C_{22}\cos^{4}\theta},\label{eq:youngs}
\end{equation}
}

\textcolor{black}{and 
\begin{equation}
\nu(\theta)=\frac{C_{12}(sin^{4}\theta+cos^{4}\theta)-(C_{11}+C_{22}-\frac{C_{11}C_{22}-C_{12}^{2}}{C_{66}})\sin^{2}\theta\cos^{2}\theta}{C_{11}\sin^{4}\theta+\left(\frac{C_{11}C_{22}-C_{12}^{2}}{C_{66}}-2C_{12}\right)\sin^{2}\theta\cos^{2}\theta+C_{22}\cos^{4}\theta}\label{eq:poisson}
\end{equation}
}

\subsection*{\textcolor{black}{Optical Properties:}}

\textcolor{black}{The frequency-dependent linear optical absorption
coefficient and the transition dipole moments (TDMs) were computed
within the Independent-Particle Approximation (IPA). Optical spectra
were first evaluated using the GGA-PBE functional \cite{perdew1996}
to establish qualitative trends, followed by calculations using more
accurate HSE06 functional \cite{Heyd2003}. While IPA neglects the
electron--hole interactions that can only be treated within many-body
approaches such as the Bethe--Salpeter equation (BSE) \cite{BSE1954},
it, however, provides a reliable description of the optical absorption
onset when combined with an improved band-gap functional. Accordingly,
the HSE06--IPA approach was adopted as a computationally efficient
and physically consistent framework for the present systems. A total
of 144 bands were included to ensure the convergence of the high-energy
optical response. The interaction of the system with incident electromagnetic
radiations of angular frequency $\omega$ is described by the complex
dielectric function, $\epsilon(\omega)=\epsilon_{Re}(\omega)+i\epsilon_{Im}(\omega)$.
The imaginary part, $\epsilon_{Im}(\omega)$, was calculated by summing
over transitions between occupied and unoccupied Kohn--Sham states,
while real part, $\epsilon_{Re}(\omega)$, was derived via the Kramers--Kronig
transformation. The optical absorption coefficient $\alpha(\omega)$
was subsequently calculated as:
\begin{equation}
\alpha(\omega)=\frac{\omega\sqrt{2}}{c}\left[\sqrt{\epsilon_{Re}^{2}(\omega)+\epsilon_{Im}^{2}(\omega)}-\epsilon_{Re}(\omega)\right]^{\frac{1}{2}}.\label{eq:absorption}
\end{equation}
}

\textcolor{black}{In addition to the macroscopic dielectric response,
TDMs were explicitly evaluated to quantify the strength of the individual
optical transitions. The TDM associated with a transition from an
initial state $i$ to a final state $f$ is defined as \cite{WANG2021} }

\textcolor{black}{
\begin{equation}
P_{i\rightarrow f}=\left\langle \psi_{f}\right|\overrightarrow{r}\left|\psi_{i}\right\rangle =\frac{i\hbar}{(E_{f}-E_{i})m}\left\langle \psi_{f}\right|\overrightarrow{p}\left|\psi_{i}\right\rangle ,\label{eq:TDM}
\end{equation}
}

\textcolor{black}{where $\psi_{i}$ ($E_{i}$) and $\psi_{f}$($E_{f}$)
are the wave functions (eigenenergies) corresponding to the initial
and final states, $m$ is the electron mass, and $\overrightarrow{r}$
and $\overrightarrow{p}$, respectively, denote the position and momentum
operators. The square of the TDMs ($P^{2}$) is proportional to the
optical transition probability, and provides quantitative feel of
the photoexcitation strength. Therefore, in this work, the $P^{2}$
values computed using the VASPKIT package \cite{WANG2021} are presented,
complementing the absorption spectra derived from the dielectric function.}

\section{\textcolor{black}{Results and discussion}}\label{sec:results-and-discussion}

\textcolor{black}{In this section, we begin with the discussion of
our calculations on the pristine s-C\textsubscript{3}N\textsubscript{6}
monolayer, followed by an investigation into the effects of uniaxial
and biaxial strains on its electronic structure and band alignment
for photocatalytic application.}

\subsection{\textcolor{black}{Pristine s-C\protect\textsubscript{3}N\protect\textsubscript{6}
Monolayer}}

\subsubsection{\textcolor{black}{Geometric structure and thermal stability}}

\begin{figure}[H]
\begin{centering}
\textcolor{black}{\includegraphics[scale=0.5]{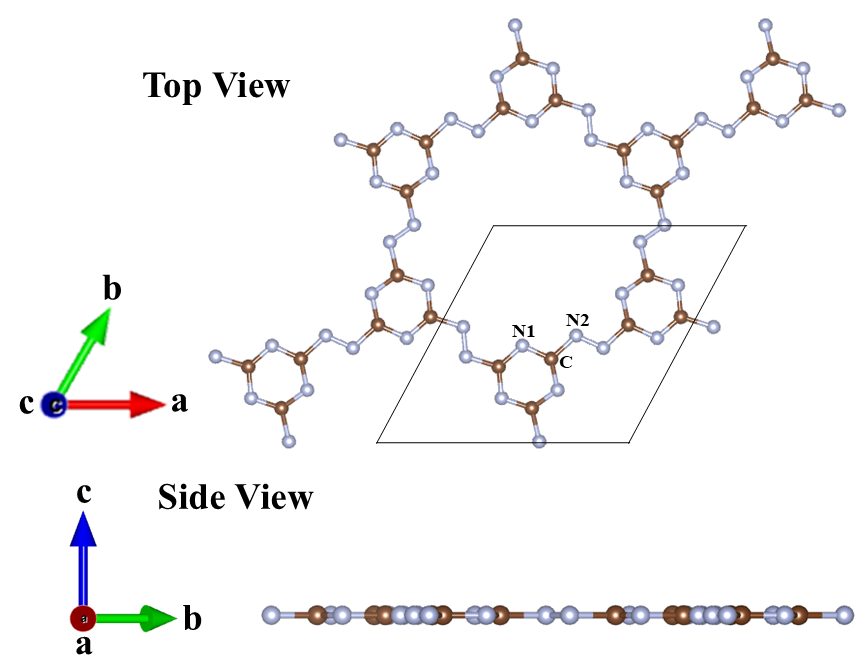}}
\par\end{centering}
\textcolor{black}{\caption{Crystal structure of $s-C_{3}N_{6}$ monolayer, where brown and gray
spheres represent the carbon and nitrogen atoms, respectively.}\label{fig:strc}
}
\end{figure}

\textcolor{black}{The crystal structure of s-C\textsubscript{3}N\textsubscript{6}
is hexagonal and belongs to the $P6/m$ space group (No. 175). The
monolayer exhibits a planar geometry composed of s-triazine units
interconnected through azo (N--N)-linkages. The unit cell contains
two s-triazine units, comprising 6 carbon and 12 nitrogen atoms, as
shown in Fig. \ref{fig:strc}. The optimized lattice constant is 10.57
$\text{\AA}$, and the calculated bond lengths for C--N1, C--N2,
N2--N2 are 1.33, 1.44, and 1.26 $\text{\AA}$, respectively, in good
agreement with previously reported values \cite{Deshpande2022}. The
absence of out-of-plane distortions confirms the planar nature of
the monolayer. The planar geometry of the s-C\textsubscript{3}N\textsubscript{6}
monolayer is advantageous as it can facilitate rapid and directional
charge mobility, which is vital for maintaining high photocatalytic
activity. To ensure the energetic stability of the considered s-C\textsubscript{3}N\textsubscript{6}
monolayer, its binding energy per atom, $E_{b}$ is calculated as:
\begin{equation}
E_{b}=\frac{1}{N_{Tot}}\left[E_{Tot}-(E_{C}N_{C}+E_{N}N_{N})\right]
\end{equation}
where $E_{Tot}$ is the total energy of the monolayer, $E_{N}$ and
$E_{C}$ are the energies of isolated nitrogen and carbon atoms, respectively,
and $N_{N}$, $N_{C}$, $N_{Tot}$ denote the number of nitrogen atoms,
carbon atoms, and total atoms in the unit cell. The calculated binding
energy of -5.59 eV/atom is close to the reported value of -5.53 eV/atom
\cite{Deshpande2022}, and indicates high energetic stability with
strong covalent bonding. }

\begin{figure}[H]
\begin{centering}
\textcolor{black}{\includegraphics[scale=0.5]{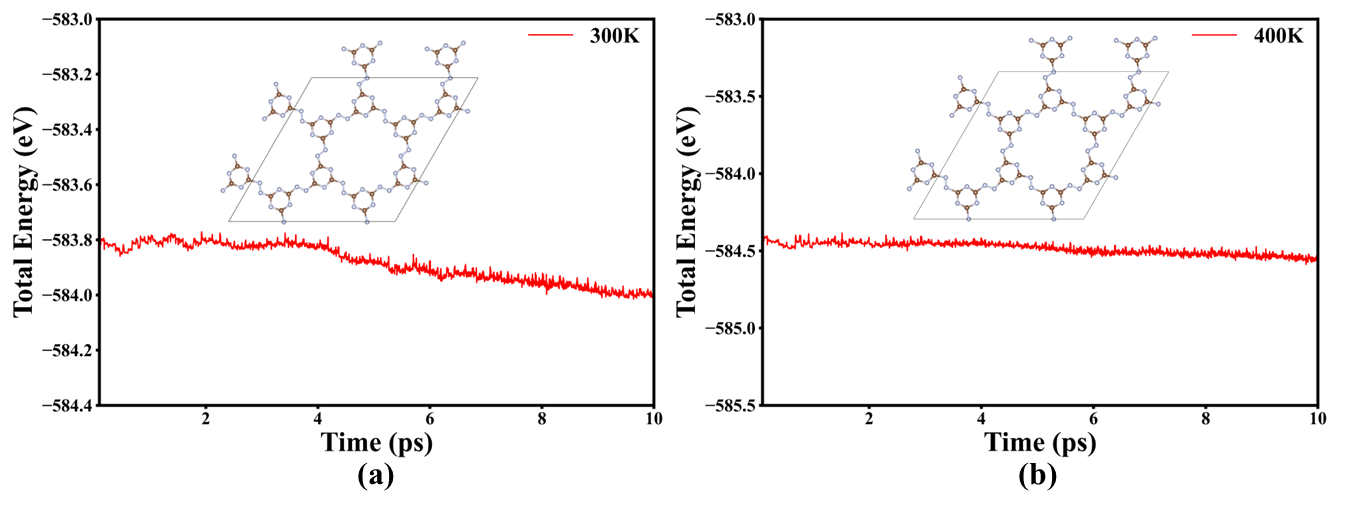}}
\par\end{centering}
\textcolor{black}{\caption{Variation of the total energy of the s-C\protect\textsubscript{3}N\protect\textsubscript{6}
monolayer as a function of time at (a) 300 K and (b) 400 K. The structural
geometries retained after 10 ps are included in the insets. }\label{fig:pristine_md}
}
\end{figure}

\begin{figure}[H]
\begin{centering}
\textcolor{black}{\includegraphics[scale=0.45]{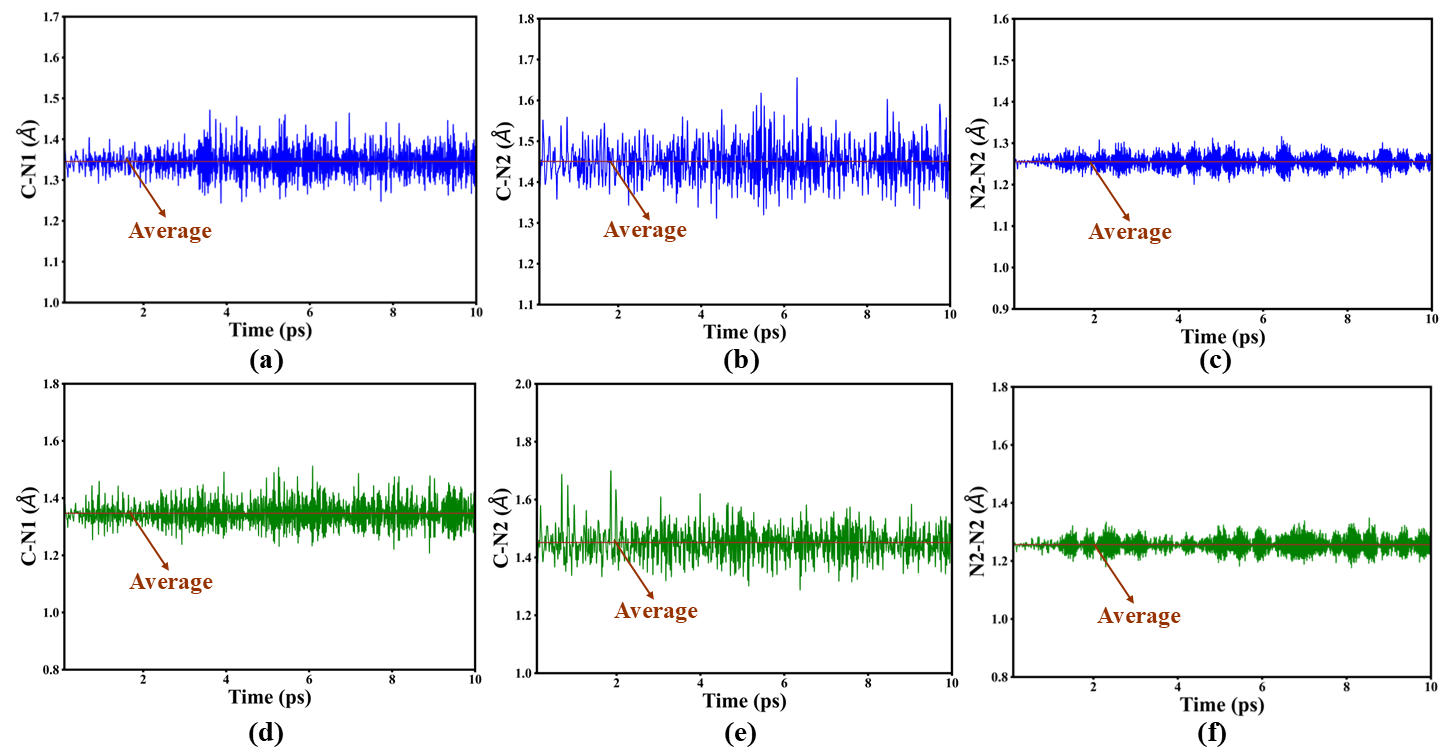}}
\par\end{centering}
\textcolor{black}{\caption{Variation of (a)\&(d) C--N1, (b)\&(e) C-N2 , and (c)\&(f) N2-N2
bond lengths of the s-C\protect\textsubscript{3}N\protect\textsubscript{6}
monolayer with time at 300 K \& 400K, respectively. }\label{fig:md300-bond}
}
\end{figure}

\textcolor{black}{Thermal stability of the s-C\textsubscript{3}N\textsubscript{6}
monolayer was further assessed using AIMD simulations performed on
a $2\times2$ supercell containing 24 carbon and 48 nitrogen atoms.
The simulations were carried out within the NVT ensemble employing
a Nosé-Hoover thermostat at temperatures of 300 K and 400 K for a
total simulation time of 10 ps with a time step of 1 fs. As shown
in Fig. \ref{fig:pristine_md}, the total energy exhibits only small
fluctuations at both temperatures, indicating thermal stability of
the monolayer. To further examine structural integrity, the time evolution
of the C--N1, C--N2, and N2--N2 bond lengths was monitored during
the simulations as presented in Fig. \ref{fig:md300-bond}. The maximum
bond-length fluctuations are limited to 2.2\% (2.9\%) for C--N1,
2.9\% (3.3\%) for C--N2, and 1.2 \% (1.9 \%) for N2--N2 bonds at
300 K (400 K). These small variations confirm the robust structural
stability of the s-C\textsubscript{3}N\textsubscript{6} monolayer
under ambient and elevated thermal conditions.}

\subsubsection{\textcolor{black}{Mechanical properties}}

\begin{figure}[H]
\begin{centering}
\textcolor{black}{\includegraphics[scale=0.5]{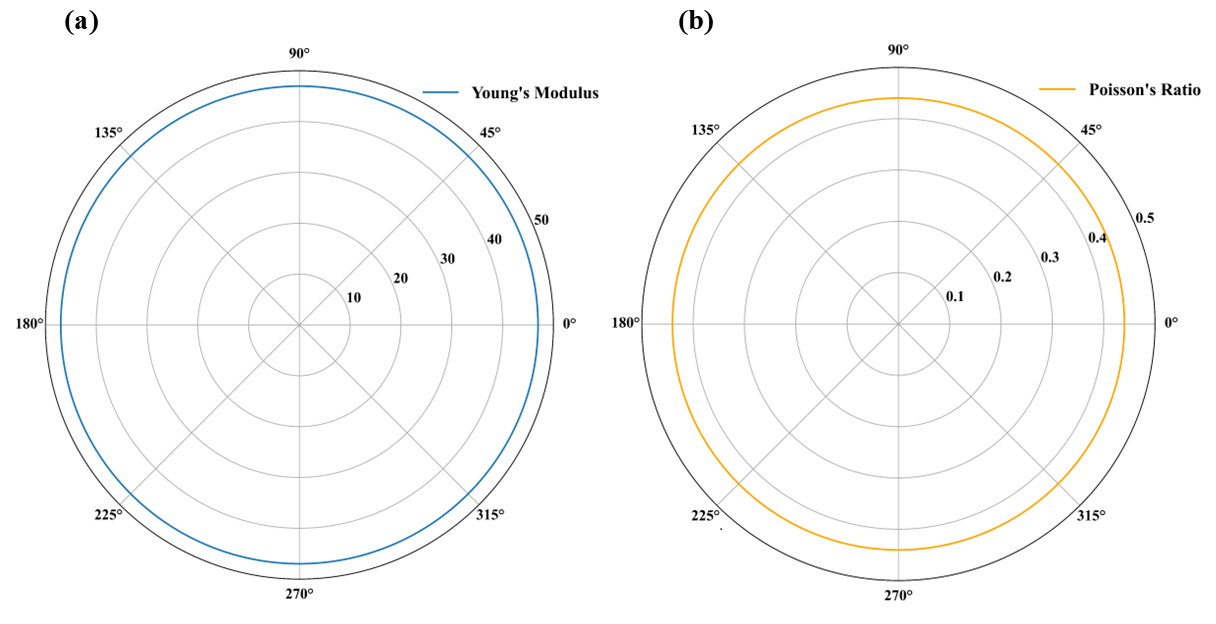}}
\par\end{centering}
\textcolor{black}{\caption{Orientation dependent (a) Young's Modulus $E(\theta)$ in the units
of Nm\protect\textsuperscript{-1} and (b) Poisson ratio $\nu(\theta)$
computed for the s-C\protect\textsubscript{3}N\protect\textsubscript{6}
monolayer.}\label{fig:Elastic}
}

\end{figure}

\textcolor{black}{The mechanical properties of the pristine $s-C_{3}N_{6}$
monolayer were examined to assess its elastic stability and suitability
for strain engineering, which is essential for practical device integration
where lattice mismatch with substrates can induce external strain.
Small in-plane strains ranging from -2\% to 2\% were applied in steps
of 0.5\%, and the corresponding total energies were calculated using
the hybrid HSE06 functional. The elastic stiffness constants were
extracted by fitting the energy--strain relations using Eq. \ref{eq:energy-strain}.
The calculated elastic constants are $C_{11}$= $C_{22}$= 58.27 Nm\textsuperscript{-1},
$C_{21}$= $C_{12}$= 25.66 Nm\textsuperscript{-1}, and $C_{66}$=
16.31 Nm\textsuperscript{-1}. These values satisfy the Born--Huang
criteria for 2D hexagonal systems, $C_{11}C_{22}-C_{12}^{2}$ > 0
and $C_{66}$> 0, confirming the elastic stability of the pristine
$s-C_{3}N_{6}$ monolayer. The shear modulus is obtained as $G=C_{66}=$16.31
Nm\textsuperscript{-1}, consistent with the symmetry relations of
a hexagonal lattice. Using the computed elastic constants, the in-plane
orientation-dependent Young\textquoteright s modulus $E(\theta)$
and Poisson's ratio $\nu(\theta)$ were determined from Eq. \ref{eq:youngs}
and Eq. \ref{eq:poisson}, respectively. The polar plots shown in
Figs. \ref{fig:Elastic}(a) and Fig. \ref{fig:Elastic}(b) indicate
that both quantities are isotropic within the plane, reflecting the
hexagonal symmetry of the crystal structure. The Young\textquoteright s
modulus is found to be 46.98 Nm\textsuperscript{-1}, while the Poisson's
ratio is 0.44, indicating moderate in-plane stiffness and elastic
flexibility. The positive value of the Poisson\textquoteright s ratio
implies lateral contraction (expansion) under tensile (compressive)
strain, which is characteristic of conventional elastic materials.
The combination of elastic stability, mechanical isotropy, and moderate
stiffness suggests that the $s-C_{3}N_{6}$ monolayer can sustain
externally applied strain without structural degradation. This supports
its feasibility for strain-engineered tuning of the electronic and
optical properties in applied settings.}

\subsubsection{\textcolor{black}{Electronic structure }}\label{subsec:Electronic-structure}

\begin{figure}[H]
\begin{centering}
\textcolor{black}{\includegraphics[scale=0.5]{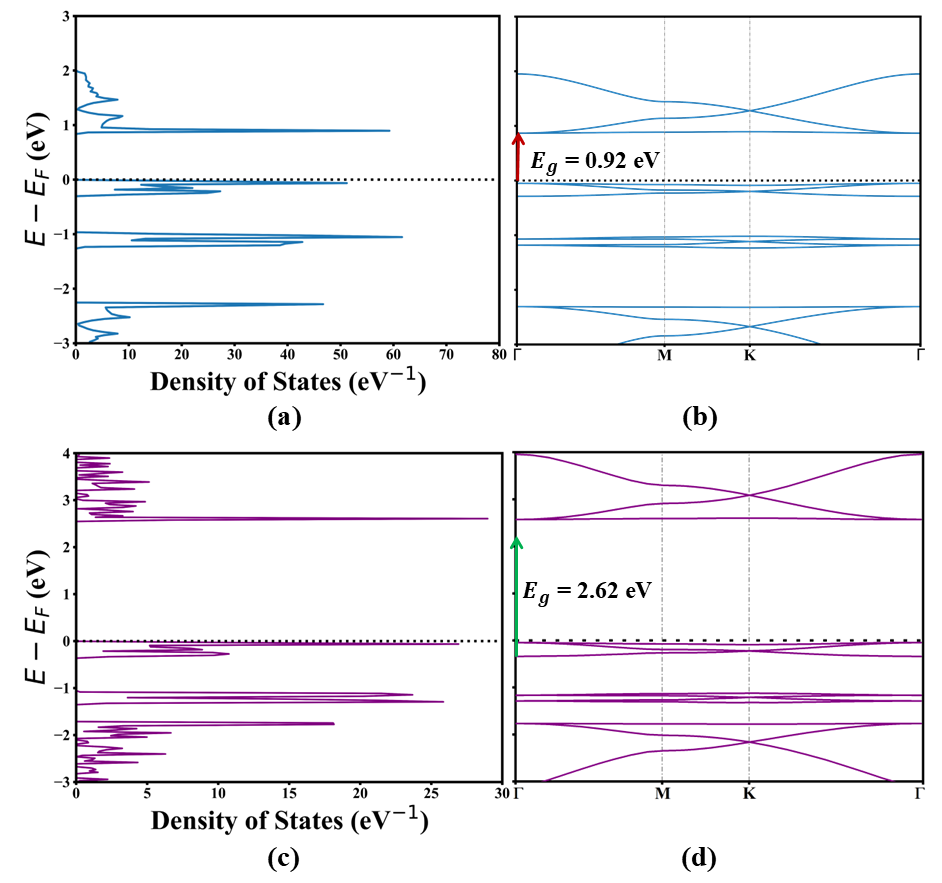}}
\par\end{centering}
\textcolor{black}{\caption{(a) \& (c) Electronic density of states, and (b) \& (d) band structures
of the s-C\protect\textsubscript{3}N\protect\textsubscript{6} monolayer
calculated using the PBE \& HSE06 functionals, respectively. The dotted
line at zero represents the Fermi level ($E_{F}$).}\label{fig:pristine-dos-band}
}
\end{figure}

\textcolor{black}{The electronic density of states (DOS) and band
structure of the pristine s-C\textsubscript{3}N\textsubscript{6}
monolayer were calculated using both the GGA-PBE and HSE06 functionals,
as shown in Fig. \ref{fig:pristine-dos-band}. The total DOS and band
dispersion are consistent with each other. At the PBE level, the monolayer
is found to be a direct-gap semiconductor with a band gap ($E_{g}$)
of 0.92 eV, with both the valence band maximum (VBM) and conduction
band minimum (CBM) located at the $\Gamma$ point (Fig. \ref{fig:pristine-dos-band}(b)).
On employing the HSE06 functional, $E_{g}$ increases to 2.62 eV (Fig.
\ref{fig:pristine-dos-band}(d)), in good agreement with previously
reported results. Importantly, the direct nature of the band gap is
preserved when moving from PBE to HSE06, indicating that the hybrid
functional primarily introduces a rigid upward shift of the conduction-band
states, thereby widening the gap, without altering the underlying
band topology. The resulting direct HSE06 band gap is desirable for
optoelectronic and photocatalytic applications as it permits the direct
photo-excitation of an electron from the valence band to the conduction
band without requiring phonon interaction to conserve momentum. This
leads to a greater quantum yield for electron-hole pair generation,
a critical factor for high photocatalytic efficiency. }

\begin{figure}[H]
\begin{centering}
\textcolor{black}{\includegraphics[scale=0.4]{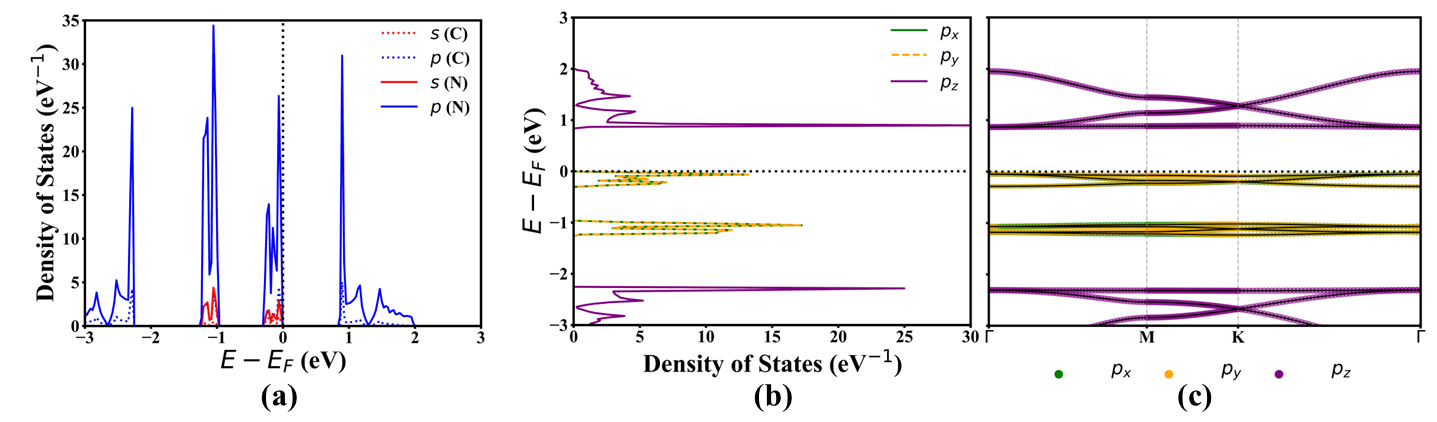}}
\par\end{centering}
\textcolor{black}{\caption{(a) $l$-decomposed partial density of states showing the contributions
of $s$ and $p$ orbitals of both carbon and nitrogen atoms, $lm$-decomposed
(b) partial density of states and (c) corresponding band structure
of the s-C\protect\textsubscript{3}N\protect\textsubscript{6} monolayer
showing the contribution of $p$ orbitals of nitrogen. The dotted
line at zero represents $E_{F}$.}\label{fig:=000020pdos}
}
\end{figure}

\textcolor{black}{To gain deeper insights into the electronic structure,
the $l$-decomposed partial density of states (PDOS) for each atom
type are also analyzed (Fig. \ref{fig:=000020pdos} (a)). It is observed
that the maximum contribution to states near the Fermi level ($E_{F}$),
including both the VBM and CBM, originates primarily from the $p$
orbitals of the nitrogen atoms. This is because the number of nitrogen
atoms in the system is twice that of the carbon atoms. Following this,
$lm$-decomposed PDOS and the projected band structure (Fig. \ref{fig:=000020pdos}
(b) and (c)) confirm that $p$ orbitals from nitrogen atoms dominate
these states. Particularly, the VBM is composed primarily of in-plane
$p_{x}$ and $p_{y}$ orbitals, which contribute to $\sigma$ bonds.
In contrast, the CBM originates from out-of-plane $p_{z}$ orbitals
of nitrogen, reflecting a dominant $\pi$ character. This composition,
where the VBM and CBM arise from distinctly oriented orbitals, can
inherently facilitate the efficient spatial separation of photo-generated
electrons and holes.}

\subsection{\textcolor{black}{Strained s-C\protect\textsubscript{3}N\protect\textsubscript{6}
monolayer}}

\begin{figure}[H]
\begin{centering}
\textcolor{black}{\includegraphics[scale=0.4]{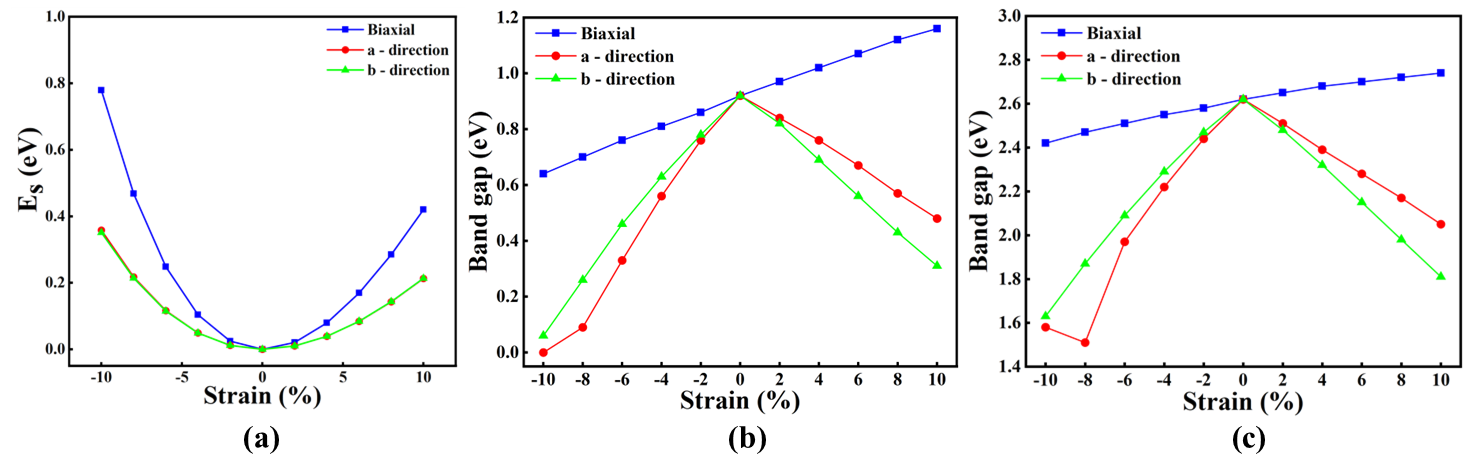}}
\par\end{centering}
\textcolor{black}{\caption{(a) Strain energy ($E_{s}$) versus strain curve, fundamental electronic
band gap variation of the s-C\protect\textsubscript{3}N\protect\textsubscript{6}
monolayer with applied strain using the (b) GGA-PBE, and (c) HSE06
functionals for all the considered strains.}\label{fig:strain-gap}
}
\end{figure}

\textcolor{black}{To engineer the electronic structure of the s-C\textsubscript{3}N\textsubscript{6}
monolayer, in-plane strain was applied, defined as $\epsilon=(a-a_{0})/a_{0}$,
where $a_{0}$ and $a$ are the lattice constants of the pristine
and strained structures, respectively. Three strain configurations
were considered within a \textpm 10\% range: (i) isotropic biaxial
strain, (ii) uniaxial strain along the $a$ direction, and (iii) uniaxial
strain along the $b$ direction. The positive (negative) values of
the strain correspond to tensile (compressive) strain. Within the
considered range of \textpm 10\% strain, the maximum external pressure
corresponds to the \textminus 10\% biaxial strain and is calculated
to be 13 GPa. This pressure is experimentally accessible using diamond
anvil cell techniques \cite{Shenghai2022}. While achieving a -10\%
biaxial strain is experimentally demanding, such magnitudes can be
realized by transferring 2D monolayers onto flexible substrates or
pre-patterned surfaces, where strong van der Waals adhesion effectively
stabilizes the lattice \cite{Metzger2010,Changgu2008,Bertolazzi2011}.
For instance, strains exceeding 10\% have been successfully demonstrated
in graphene \cite{Changgu2008} and MoS\textsubscript{2} monolayer
\cite{Bertolazzi2011} using these substrate-mediated techniques.
To address potential issues such as buckling or delamination, experimentalists
can employ thin-film encapsulation (e.g. using h-BN) or utilize the
thermal expansion mismatch between the monolayer and the substrate
to \textquotedbl trap\textquotedbl{} the compressive state. These
methods provide a mechanical clamping effect that suppresses out-of-plane
distortions even at high loads. To ensure that all applied strains
remain within the elastic regime, the strain energy $E_{s}=E_{strained}-E_{unstrained}$
was evaluated. As shown in Fig. \ref{fig:strain-gap}(a), the quadratic
dependence of $E_{s}$ on strain confirms elastic and reversible deformation
for all strain modes. The nearly identical strain energies for uniaxial
loading along the $a$ and $b$ directions further reflect the intrinsic
in-plane isotropy of the monolayer. }

\subsubsection{\textcolor{black}{Strain-dependent electronic structure }}

\begin{figure}[H]
\begin{centering}
\textcolor{black}{\includegraphics[scale=0.45]{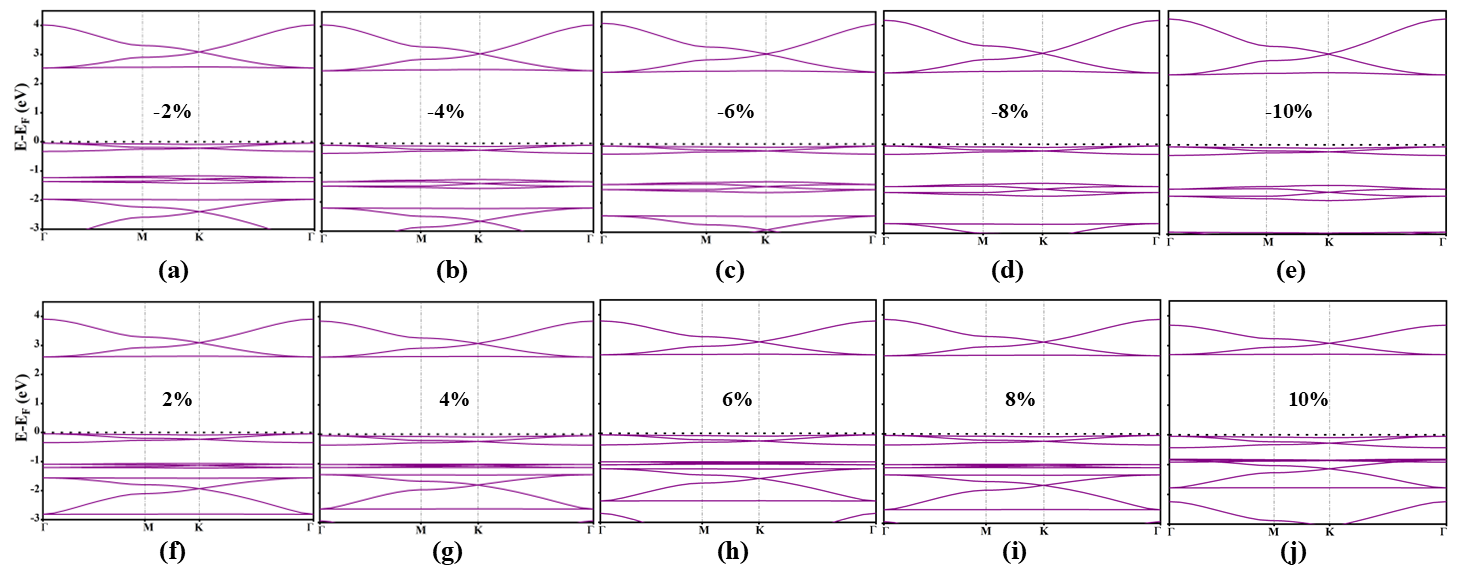}}
\par\end{centering}
\textcolor{black}{\caption{Calculated band structures using the HSE06 functional for the s-C\protect\textsubscript{3}N\protect\textsubscript{6}
monolayer subject to biaxial strains of (a) -2\%, (b) -4\%, (c) -6\%,
(d) -8\%, (e) -10\%, (f) 2\%, (g) 4\%, (h) 6\%, (i) 8\%, (j) 10\%. }\label{fig:HSE06-band}
}

\end{figure}

\textcolor{black}{The electronic band structures under strain were
calculated using both the GGA-PBE and HSE06 functionals. As shown
in Figs. \ref{fig:strain-gap}(b) and (c), both functionals predict
qualitatively similar strain-dependent trends in the band gap $E_{g}$,
with HSE06 yielding systematically larger values. Under biaxial strain,
$E_{g}$ increases monotonically with tensile strain and decreases
under compression. In contrast, uniaxial strain along either $a$
or $b$ directions leads to a reduction of $E_{g}$ under both compressive
and tensile loading. For biaxial strain, both the GGA-PBE (Fig. S1
of the Supporting Information (SI)) and HSE06 band structures (Fig.
\ref{fig:HSE06-band}) indicate that strain primarily induces a rigid
shift of the band edges without altering the overall band topology.
In particular, the direct band gap at the $\Gamma$ point is preserved
throughout the entire biaxial strain range. This behavior is attributed
to the retention of hexagonal symmetry ($P6/m$) under isotropic deformation.
In contrast, uniaxial strain lowers the crystal symmetry from hexagonal
$P6/m$ to monoclinic $P2/m$, leading to anisotropic modifications
of the electronic states. This symmetry reduction leads to anisotropic
modifications of the electronic bands under uniaxial strain as shown
in Figs. S2--S5 of the SI for both the GGA-PBE and HSE06 calculations.
For most uniaxial strain configurations, the direct band gap remains
located at the $\Gamma$ point. However, under \textminus 8\% uniaxial
strain applied along the $a$ direction, a shift of the direct band
gap to the M point is observed using the HSE06 functional. Since the
HSE06 functional corrects the band-gap underestimation inherent to
GGA-PBE while preserving the qualitative strain-induced trends, it
was employed exclusively for subsequent evaluations of band-edge alignment
and photocatalytic performance in the strained s-C\textsubscript{3}N\textsubscript{6}
monolayer.}

\subsubsection{\textcolor{black}{Thermodynamic criteria for photocatalysis: band
alignment}}

\textcolor{black}{The photocatalytic water splitting is fundamentally
driven by the photo-excitation of charge carriers, followed by their
migration to the catalyst surface to initiate redox reactions. This
overall process is governed by two half-reactions: the reduction of
protons to generate hydrogen, i.e., the hydrogen evolution reaction
(HER), and the oxidation of water to produce oxygen leading to the
oxygen evolution reaction (OER). These reduction and oxidation reactions
are represented as:}

\textcolor{black}{
\begin{equation}
2H^{+}+2e^{-}\rightarrow H_{2}\label{eq:red-1}
\end{equation}
}

\textcolor{black}{
\begin{equation}
H_{2}O+2h^{+}\rightarrow\frac{1}{2}O_{2}+2H^{+}\label{eq:oxi-1}
\end{equation}
For spontaneous overall water splitting, a semiconductor must satisfy
two criteria: (i) a band gap larger than 1.23 eV, and (ii) appropriate
alignment of the band edges relative to the redox potentials of water.
At pH = 0, the reduction $(E_{red})$ and oxidation $(E_{ox})$ potentials
are \textminus 4.44 eV and \textminus 5.67 eV, respectively, relative
to the vacuum level \cite{Singh2015}. The first thermodynamic requirement
comes from the standard Gibbs free energy change ($\Delta G$ = 237
kJ/mol) needed to drive the endergonic water splitting reaction \cite{Kudo2009}.
The pristine s-C\textsubscript{3}N\textsubscript{6} monolayer satisfies
the first criterion with an HSE06 band gap of 2.62 eV but fails to
meet the second due to an unfavorable conduction band position. Specifically,
while the VBM at \textminus 7.515 eV is suitable for OER, the CBM
at \textminus 4.897 eV lies below the reduction potential, preventing
HER.}

\begin{figure}[H]
\begin{centering}
\textcolor{black}{\includegraphics[scale=0.5]{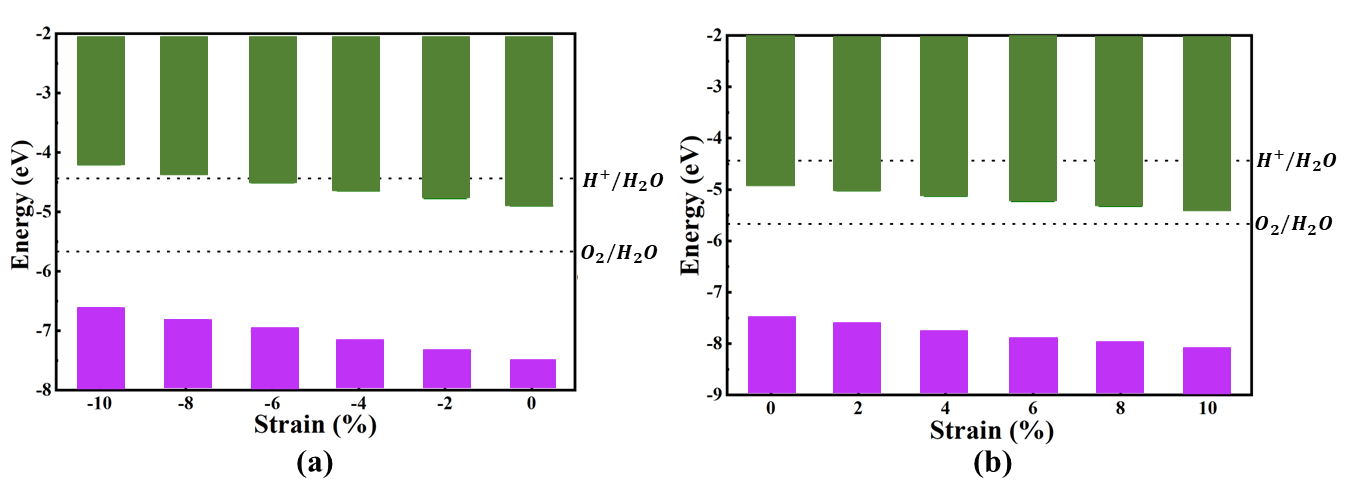}}
\par\end{centering}
\textcolor{black}{\caption{CBM and VBM positions of the s-C\protect\textsubscript{3}N\protect\textsubscript{6}
monolayer under the influence of (a) compressive, and (b) tensile
biaxial strains. Redox potentials are shown by dotted lines.}\label{fig:Band-edges}
}
\end{figure}

\textcolor{black}{Strain engineering provides an effective route to
overcome this limitation. The strain-dependent band-edge positions
($E_{CBM}$ and $E_{VBM}$) relative to the vacuum level are summarized
in Table \ref{tab:8_band-edge-table}. The band-edge positions for
biaxial strain with respect to the redox potentials are illustrated
in the band alignment diagram of Fig. \ref{fig:Band-edges}. Under
compressive biaxial strain (Fig. \ref{fig:Band-edges}(a)), the CBM
shifts upward (becomes less negative), while the VBM also shifts upward
but remains sufficiently deep to sustain oxidation. Notably, at \textminus 8\%
and \textminus 10\% biaxial strain, the CBM crosses the reduction
potential, yielding a positive reducing power $\Delta_{R}=E_{CBM}-E_{red}$,
while the oxidizing power $\Delta_{O}=E_{ox}-E_{VBM}$ also remains
positive. Therefore, these strain configurations satisfy thermodynamic
requirements for both HER and OER and thus can facilitate overall
water splitting. In contrast, tensile biaxial strain (Fig. \ref{fig:Band-edges}(b))
and all uniaxial strain configurations (see Fig. S6 of the SI for
band alignment) shift the CBM to more negative energies relative to
the reduction potential compared to the pristine monolayer, thereby
failing to satisfy the band-alignment requirement for HER. It is further
noted that the CBM and VBM positions for the \textminus 8\% uniaxial
strain applied along the $a$ direction deviate from the monotonic
trend observed over the \textminus 2\% to \textminus 10\% strain range.
This deviation correlates with the relocation of the direct band gap
from the $\Gamma$ point to the M point under this specific strain
condition, leading to a modified distribution of the band-edge states
in momentum space. Our analysis confirms that only the compressive
biaxial strains of -8\% and -10\% are effective in achieving the required
band alignment, transforming the s-C\textsubscript{3}N\textsubscript{6}
monolayer into a viable photocatalyst for spontaneous overall water
splitting. }

\begin{table}[H]
\textcolor{black}{\caption{Band edge positions with respect to the vacuum level, $E_{VBM}$
and $E_{CBM}$ in eV for the considered strained s-C\protect\textsubscript{3}N\protect\textsubscript{6}
monolayers calculated using the HSE06 functional.}\label{tab:8_band-edge-table}
}
\centering{}%
\begin{tabular}{ccccccc}
\toprule 
\multirow{2}{*}{\textcolor{black}{Strain}} & \multicolumn{2}{c}{\textcolor{black}{Biaxial}} & \multicolumn{2}{c}{\textcolor{black}{Uniaxial along $a$}} & \multicolumn{2}{c}{\textcolor{black}{Uniaxial along $b$}}\tabularnewline
\cmidrule{2-7}
 & \textcolor{black}{$E_{VBM}$} & \textcolor{black}{$E_{CBM}$} & \textcolor{black}{$E_{VBM}$} & \textcolor{black}{$E_{CBM}$} & \textcolor{black}{$E_{VBM}$} & \textcolor{black}{$E_{CBM}$}\tabularnewline
\midrule
\midrule 
\textcolor{black}{-10 \%} & \textcolor{black}{-6.623} & \textcolor{black}{-4.206} & \textcolor{black}{6.733} & \textcolor{black}{-5.155} & \textcolor{black}{-6.750} & \textcolor{black}{-5.119}\tabularnewline
\midrule 
\textcolor{black}{-8 \%} & \textcolor{black}{-6.824} & \textcolor{black}{-4.359} & \textcolor{black}{-6.197} & \textcolor{black}{-4.489} & \textcolor{black}{-6.943} & \textcolor{black}{-5.076}\tabularnewline
\midrule 
\textcolor{black}{-6 \%} & \textcolor{black}{-7.015} & \textcolor{black}{-4.506} & \textcolor{black}{-7.037} & \textcolor{black}{-5.063} & \textcolor{black}{-7.122} & \textcolor{black}{-5.033}\tabularnewline
\midrule 
\textcolor{black}{-4 \%} & \textcolor{black}{-7.193} & \textcolor{black}{-4.645} & \textcolor{black}{-7.230} & \textcolor{black}{-5.006} & \textcolor{black}{-7.279} & \textcolor{black}{-4.989}\tabularnewline
\midrule 
\textcolor{black}{-2 \%} & \textcolor{black}{-7.362} & \textcolor{black}{-4.777} & \textcolor{black}{-7.392} & \textcolor{black}{-4.948} & \textcolor{black}{-7.409} & \textcolor{black}{-4.94}\tabularnewline
\midrule 
\textcolor{black}{0 \%} & \textcolor{black}{-7.515} & \textcolor{black}{-4.897} & \textcolor{black}{-7.515} & \textcolor{black}{-4.897} & \textcolor{black}{-7.515} & \textcolor{black}{-4.897}\tabularnewline
\midrule 
\textcolor{black}{2 \%} & \textcolor{black}{-7.668} & \textcolor{black}{-5.002} & \textcolor{black}{-7.565} & \textcolor{black}{-5.060} & \textcolor{black}{-7.547} & \textcolor{black}{-5.064}\tabularnewline
\midrule 
\textcolor{black}{4 \%} & \textcolor{black}{-7.803} & \textcolor{black}{-5.127} & \textcolor{black}{-7.598} & \textcolor{black}{-5.204} & \textcolor{black}{-7.542} & \textcolor{black}{-5.221}\tabularnewline
\midrule 
\textcolor{black}{6 \%} & \textcolor{black}{-7.936} & \textcolor{black}{-5.231} & \textcolor{black}{-7.629} & \textcolor{black}{-5.345} & \textcolor{black}{-7.528} & \textcolor{black}{-5.378}\tabularnewline
\midrule 
\textcolor{black}{8 \%} & \textcolor{black}{-8.047} & \textcolor{black}{-5.323} & \textcolor{black}{-7.636} & \textcolor{black}{-5.468} & \textcolor{black}{-7.509} & \textcolor{black}{-5.532}\tabularnewline
\midrule 
\textcolor{black}{10 \%} & \textcolor{black}{-8.128} & \textcolor{black}{-5.385} & \textcolor{black}{-7.633} & \textcolor{black}{-5.584} & \textcolor{black}{-7.486} & \textcolor{black}{-5.678}\tabularnewline
\bottomrule
\end{tabular}
\end{table}

\subsubsection{\textcolor{black}{Mechanical and thermal stability of the photocatalytically
active configurations}}

\textcolor{black}{Having identified compressive biaxial strains of
\textminus 8\% and \textminus 10\% as the only configurations that
satisfy the band-alignment requirements for overall water splitting,
we next examine their mechanical and thermal stability to assess experimental
feasibility.}

\paragraph{\textcolor{black}{Mechanical stability:}}

\textcolor{black}{The mechanical robustness of the photocatalytically
active configurations was evaluated by calculating the elastic stiffness
constants using the same methodology as for the pristine monolayer.
Because isotropic biaxial strain preserves the hexagonal $P6/m$ symmetry,
the elastic response is fully described by the in-plane constants
$C_{11}$ and $C_{12}$. Both strained configurations exhibit pronounced
strain-stiffening behavior, characterized by a significant increase
in elastic moduli under compression. For \textminus 8\% (\textminus 10\%)
biaxial strain, the elastic constants increase to $C_{11}=106.59(123.86)$
Nm\textsuperscript{-1}, $C_{12}=61.42(74.73)$ Nm\textsuperscript{-1},
yielding $C_{66}=22.56(24.56)$ Nm\textsuperscript{-1}. This enhancement
reflects the anharmonic nature of the interatomic interactions, as
reduced bond lengths drive the system into a more strongly repulsive
region of the potential energy surface. Importantly, all elastic constants
satisfy the Born--Huang stability criteria, confirming that the s-C\textsubscript{3}N\textsubscript{6}
monolayer remains mechanically stable even under substantial compressive
strain. Further, the enhanced stiffness mitigates concerns regarding
out-of-plane buckling. This reinforces the feasibility of utilizing
high-strain states to achieve optimal band alignment for solar-to-hydrogen
conversion. This robustness is further corroborated by the in-plane
Young's modulus and Poisson's ratio, which remain isotropic and are
enhanced to $E=71.19(79.00)$ Nm\textsuperscript{-1} and $\nu=0.58(0.60)$
for the -8\% (-10\%) strained systems, respectively.}

\paragraph{\textcolor{black}{Thermal stability:}}

\textcolor{black}{Thermal stability was assessed through AIMD simulations
performed at 300 K and 400 K for both \textminus 8\% and \textminus 10\%
biaxially strained configurations. As shown in Fig. S7 of the SI,
the total energies fluctuate narrowly around their equilibrium values,
indicating stable thermal behavior at ambient and elevated temperatures.
Structural integrity was further quantified by monitoring the time
evolution of the C--N1, C--N2, and N2--N2 bond lengths. For \textminus 8\%
strain, the maximum bond-length variations are limited to 2.7\%, 2.0\%,
and 2.4\% at 300 K, increasing slightly to 3.2\%, 2.3\%, and 2.7\%
at 400 K. For \textminus 10\% strain, the corresponding variations
remain below 3.0\% (2.8\%), 5.9\% (5.5\%), and 2.4\% (3.0\%) at 300
K (400 K). Throughout the simulation time of 10 ps, no bond breaking
or significant structural distortion is observed.}

\textcolor{black}{Collectively, the strain-stiffening mechanical response
and robust thermal stability confirm that the compressively strained
s-C\textsubscript{3}N\textsubscript{6} monolayer can sustain the
deformation required to access its photocatalytically active phase.
These results establish the proposed strain modulation as not only
theoretically viable but also experimentally realistic.}

\subsubsection{\textcolor{black}{Effective masses of the charge carriers }}

\textcolor{black}{The effective masses of charge carriers in a semiconductor
play a crucial role in determining their mobility and, consequently,
the efficiency of photoinduced processes such as photocatalysis. Therefore
effective masses of both electrons ($m_{e}^{*}$) and holes ($m_{h}^{*}$)
were evaluated for the biaxially strained structures using a parabolic
approximation around the CBM and VBM, respectively. The effective-mass
tensor was computed as:
\begin{equation}
m_{ij}^{*}(\boldsymbol{k})=\hbar^{2}\left(\frac{\partial^{2}(E(\boldsymbol{k}))}{\partial k_{i}k_{j}}\right)_{\boldsymbol{k}=\boldsymbol{K_{0}}}^{-1}m_{0}
\end{equation}
}

\textcolor{black}{where $\hbar$ is the reduced Planck constant, }\textbf{\textcolor{black}{$\boldsymbol{k}$}}\textcolor{black}{{}
is the wave vector, $E(\boldsymbol{k})$ is the Kohn-Sham eigenvalue,
}\textbf{\textcolor{black}{$\boldsymbol{K_{0}}$}}\textcolor{black}{{}
denotes the band extremum, and $m_{0}$ is the free electron mass.
Owing to the in-plane isotropy of the $s-C_{3}N_{6}$ monolayer, a
single in-plane effective mass characterizes both electrons and holes,
as listed in Table \ref{tab:Effective-mass}. The calculated effective
masses are relatively large for both pristine and strained systems,
reflecting the weak band dispersion associated with the localized
$p$-orbital character of the triazine framework. Such behavior is
typical of triazine-based covalent organic frameworks and covalent
triazine frameworks, where orbital localization leads to flat bands
and heavy carriers \cite{Gutzler2016,Raptakis2022}. For example,
Gutzler }\textit{\textcolor{black}{et al.}}\textcolor{black}{{} \cite{Gutzler2016}
reported 42.83$m_{0}$(planar) and 22.35$m_{0}$ (buckled) for porous
graphene systems containing 8 phenyl rings per unit cell. Despite
these large effective masses, these materials remain promising for
photocatalysis due to their favorable band alignment and strong light
absorption \cite{Meier2019,Xie2022}.}

\begin{table}
\textcolor{black}{\caption{Calculated effective masses of electrons $m_{e}^{*}$ and holes $m_{h}^{*}$
in term of the free electron mass $m_{0}$ for the biaxailly strained
$s-C_{3}N_{6}$ monolayer.}\label{tab:Effective-mass}
}
\centering{}%
\begin{tabular}{ccc}
\toprule 
\textcolor{black}{Strain} & \textcolor{black}{$m_{h}^{*}$($m_{0}$)} & \textcolor{black}{$m_{e}^{*}$($m_{0}$)}\tabularnewline
\midrule
\midrule 
\textcolor{black}{-10} & \textcolor{black}{19.90} & \textcolor{black}{11.17}\tabularnewline
\midrule 
\textcolor{black}{-8} & \textcolor{black}{19.59} & \textcolor{black}{12.78}\tabularnewline
\midrule 
\textcolor{black}{-6} & \textcolor{black}{19.24} & \textcolor{black}{14.59}\tabularnewline
\midrule 
\textcolor{black}{-4} & \textcolor{black}{18.78} & \textcolor{black}{16.68}\tabularnewline
\midrule 
\textcolor{black}{-2} & \textcolor{black}{18.18} & \textcolor{black}{19.07}\tabularnewline
\midrule 
\textcolor{black}{0} & \textcolor{black}{17.44} & \textcolor{black}{21.81}\tabularnewline
\midrule 
\textcolor{black}{2} & \textcolor{black}{16.58} & \textcolor{black}{25.13}\tabularnewline
\midrule 
\textcolor{black}{4} & \textcolor{black}{15.61} & \textcolor{black}{28.54}\tabularnewline
\midrule 
\textcolor{black}{6} & \textcolor{black}{14.64} & \textcolor{black}{32.93}\tabularnewline
\midrule 
\textcolor{black}{8} & \textcolor{black}{13.60} & \textcolor{black}{37.35}\tabularnewline
\midrule 
\textcolor{black}{10} & \textcolor{black}{12.64} & \textcolor{black}{42.87}\tabularnewline
\bottomrule
\end{tabular}
\end{table}

\textcolor{black}{Across the investigated range of -10\% to 10\% biaxial
strain, $m_{e}^{*}$ and $m_{h}^{*}$ exhibit distinct trends. The
former increases due to reduced overlap of $p_{z}$ orbitals, whereas,
later $m_{h}^{*}$ decreases due to the transition from a highly localized
state of nitrogen lone pairs under compression to a more delocalized
electronic environment under tensile strain. Under compressive biaxial
strain, $m_{e}^{*}$ decreases significantly from 21.81$m_{0}$ in
the pristine state to 11.17$m_{0}$ at -10\% strain, indicating enhanced
electron mobility. This reduction is particularly relevant for the
photocatalytically active strain range (\textminus 8\% to \textminus 10\%).
In contrast, $m_{h}^{*}$ shows only a modest increase under compression,
from 17.44$m_{0}$ to 19.90$m_{0}$. Although the large effective
masses imply limited carrier mobility, the flat-band nature of the
electronic structure results in a high density of states near the
band edges. This enhances carrier availability at the relevant redox
potentials. Furthermore, the symmetry-forbidden nature of the fundamental
band gap (discussed below in the Sec \ref{subsec:Optical-response})
suppresses radiative recombination, leading to extended carrier lifetimes.
These factors collectively mitigate the impact of reduced mobility,
and support efficient charge utilization in photocatalytic processes.}

\subsubsection{\textcolor{black}{Optical response}}\label{subsec:Optical-response}

\begin{figure}[H]
\begin{centering}
\textcolor{black}{\includegraphics[scale=0.5]{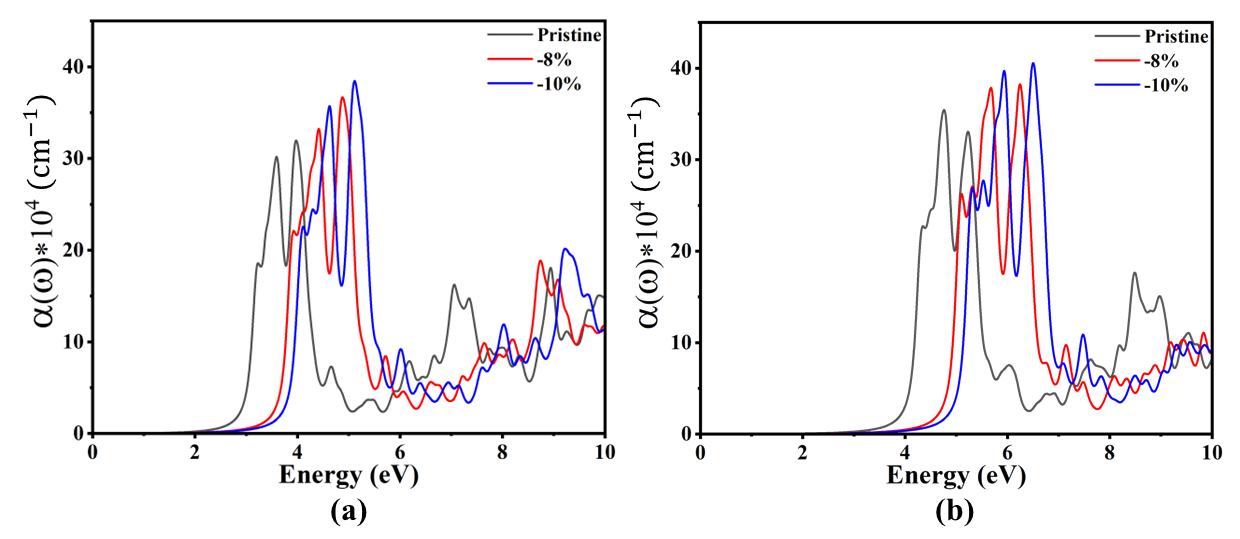}}
\par\end{centering}
\textcolor{black}{\caption{Optical absorption coefficient $\alpha(\omega)$ of the pristine
and strained s-C\protect\textsubscript{3}N\protect\textsubscript{6}
monolayers, as a function of the energy of the incident photon, calculated
at the (a) GGA-PBE, and (b) HSE06 level. }\label{fig:Optical-absorption}
}

\end{figure}

\textcolor{black}{To assess the optical behavior relevant to the photocatalytic
performance, we investigated the frequency-dependent optical response
of the s-C\textsubscript{3}N\textsubscript{6} monolayer, and its
evolution under strain. The analysis is primarily restricted to the
pristine structure, and the biaxially compressed \textminus 8\% and
\textminus 10\% strain configurations, as only these strained states
satisfy the band-alignment criteria for the overall water splitting
discussed above. The complex dielectric function $\epsilon(\omega)$
was calculated at the IPA level using both the GGA-PBE and HSE06 functionals,
and the corresponding absorption coefficients $\alpha(\omega)$ were
obtained from Eq. \ref{eq:absorption}. Both the imaginary and real
parts of the $\epsilon(\omega)$ are presented in Fig. S8 of the SI.
The absorption spectra for the in-plane polarized light are shown
in Fig. \ref{fig:Optical-absorption}. Two key features emerge from
the calculated spectra of all the three considered configurations.
First, the onset of optical absorption occurs at energies significantly
higher than the fundamental electronic band gap $E_{g}$, with the
first pronounced absorption peak appearing at an energy $E_{I}$ (see
Table \ref{tab:optical} for the numerical values). Second, the application
of biaxial compressive strain leads to a systematic blue shift of
$E_{I}$, despite the fact that $E_{g}$ decreases under compression.
Both observations are atypical for conventional semiconductors and
indicate that optical selection rules play a dominant role in governing
the absorption behavior of the s-C\textsubscript{3}N\textsubscript{6}
monolayer.}

\begin{table}[H]

\textcolor{black}{\caption{ The location of the first absorption peak position $E_{I}$, and
the values of the fundamental electronic band gap $E_{g}$, obtained
for the selected strained configurations of s-C\protect\textsubscript{3}N\protect\textsubscript{6}
monolayer under consideration. }\label{tab:optical}
}
\centering{}%
\begin{tabular}{ccccc}
\toprule 
\multirow{2}{*}{\textcolor{black}{Strain}} & \multicolumn{2}{c}{\textcolor{black}{GGA-PBE}} & \multicolumn{2}{c}{\textcolor{black}{HSE06}}\tabularnewline
\cmidrule{2-5}
 & \textcolor{black}{$E_{I}$(eV)} & \textcolor{black}{$E_{g}$(eV)} & \textcolor{black}{$E_{I}$(eV)} & \textcolor{black}{$E_{g}$(eV)}\tabularnewline
\midrule
\midrule 
\textcolor{black}{0 \%} & \textcolor{black}{3.2} & \textcolor{black}{0.92} & \textcolor{black}{4.3} & \textcolor{black}{2.62}\tabularnewline
\midrule 
\textcolor{black}{-8 \%} & \textcolor{black}{3.9} & \textcolor{black}{0.70} & \textcolor{black}{5.1} & \textcolor{black}{2.47}\tabularnewline
\midrule 
\textcolor{black}{-10 \%} & \textcolor{black}{4.1} & \textcolor{black}{0.64} & \textcolor{black}{5.3} & \textcolor{black}{2.42}\tabularnewline
\bottomrule
\end{tabular}
\end{table}

\begin{figure}[H]
\begin{centering}
\textcolor{black}{\includegraphics[scale=0.45]{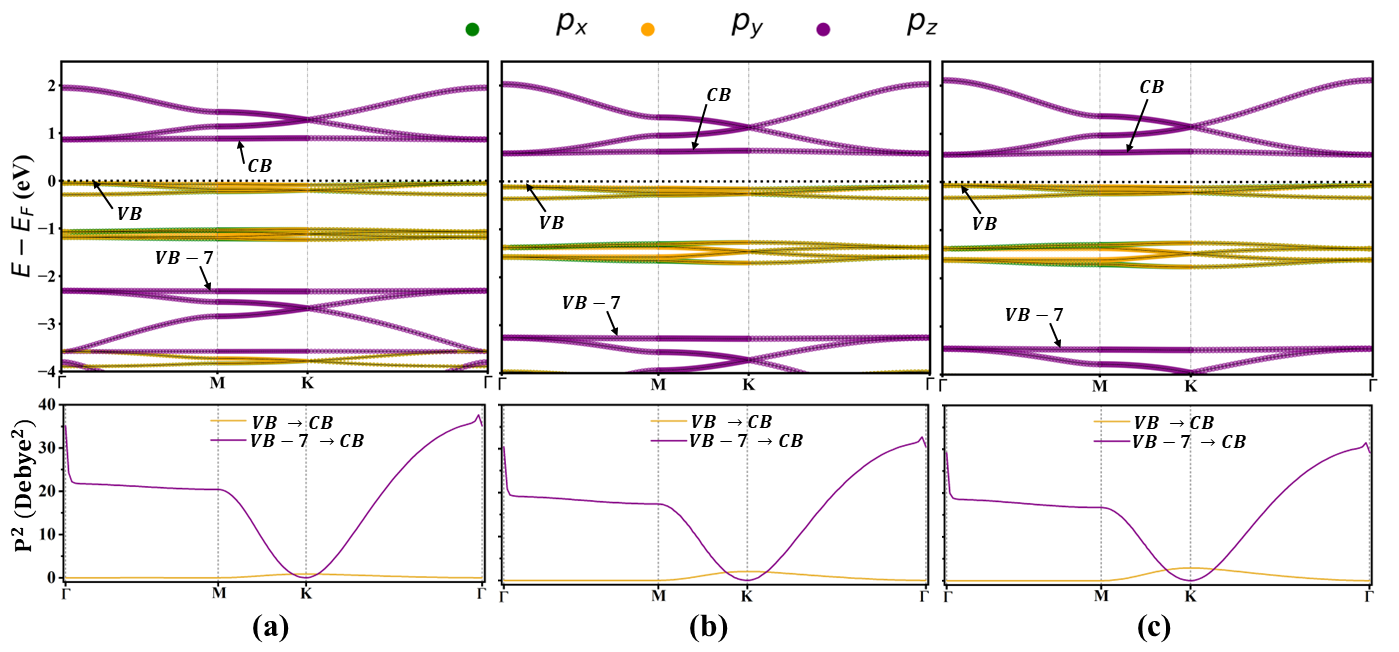}}
\par\end{centering}
\textcolor{black}{\caption{(Top row) $lm-$Projected electronic band structures calculated at
the PBE level, highlighting the contribution of nitrogen atoms' $p$-orbitals.
(Bottom row) Calculated squared transition dipole moments ($P^{2}$)
along the high-symmetry $k$-path for the two transitions, denoted
by $VB\rightarrow CB$ and $VB-7\rightarrow CB$. Panels (a), (b),
and (c) correspond to the pristine, -8\% biaxial strain, and -10\%
biaxial strain configurations of the s-C\protect\textsubscript{3}N\protect\textsubscript{6}
monolayer, respectively.}\label{fig:TDM-plots}
}
\end{figure}

\textcolor{black}{To elucidate the delayed absorption onset, we analyzed
the orbital character of the band states at the PBE level (see Fig.
\ref{fig:TDM-plots}) and evaluated the TDMs. As discussed previously,
the topmost valence band (VB) is primarily composed of in-plane $p_{x}$
and $p_{y}$ orbitals of nitrogen, while the lowest conduction band
(CB) is dominated by its out-of-plane $p_{z}$ orbitals. As a result,
the fundamental $VBM\rightarrow CBM$ transition (or $VB\rightarrow CB$
at any k-point) corresponds to a $\sigma\rightarrow\pi^{*}$ excitation,
which turns out to be symmetry forbidden, as confirmed by the vanishing
TDM (Eq. \ref{eq:TDM}) across the Brillouin zone. This conclusion
is directly confirmed by the calculated square of TDMs ($P^{2}$)
presented in Fig. \ref{fig:TDM-plots}, which are essentially zero
for the $VB\rightarrow CB$. Consequently, the electronic band gap
$E_{g}$, is optically inactive. Our analysis shows that the first
symmetry-allowed optical transition instead involves deeper valence
states, with the first such state being the seventh valence band below
the VB (denoted as VB--7), which has a dominant nitrogen $p_{z}$
character, leading to a $\pi\rightarrow\pi^{*}$($VB-7\rightarrow CB$)
excitation. This transition exhibits significant dipole strength ($P^{2}$),
with the strongest contribution at the $\Gamma$ point where the energy
separation is minimal. Therefore, the first absorption peak (Fig.
\ref{fig:Optical-absorption}) corresponds to the $VB\lyxmathsym{\textendash}7\rightarrow CB$
transition, and its position $E_{I}$ is determined by the minimum
gap between these states. The anomalous blue shift of $E_{I}$ under
biaxial compression follows naturally from this picture. Although
compression slightly reduces $E_{g}$ by lowering the CB, the forbidden
nature of the $VB\rightarrow CB$ transition remains unchanged. In
contrast, the VB--7 state shifts more strongly to lower energies
than the CB, increasing the $VB\lyxmathsym{\textendash}7\rightarrow CB$
separation. This leads to a higher-energy onset of the first allowed
transition. The same mechanism governs the absorption behavior obtained
at the HSE06 level of calculations. }

\textcolor{black}{Further, to remain consistent with the HSE06-based
band-structure and band-alignment analysis presented previously, the
optical absorption onset is also discussed at the HSE06 level. This
is important because of the well-known tendency of the GGA-PBE approach
to underestimate the band gaps. At the HSE06 level, the absorption
onset for both pristine and strained structures lies in the ultraviolet
region (see Fig. \ref{fig:Optical-absorption}). While this implies
transparency in the visible range, it does not preclude its photocatalytic
activity. Several established photocatalysts, including TiO\textsubscript{2},
SrTiO\textsubscript{3}, and Ga\textsubscript{2}O\textsubscript{3}exhibit
similar ultraviolet absorption yet remain effective due to suitable
band alignment and chemical stability \cite{FUJISHIMA1972,SHEN2025,PATIAL2020}.
Accordingly, the \textminus 8\% and \textminus 10\% biaxially strained
$s-C_{3}N_{6}$ monolayer can function as an ultraviolet-active photocatalyst.}

\textcolor{black}{While our primary analysis focuses on the photocatalytically
active biaxial configurations, we also examined the optical response
under uniaxial deformation. Specifically, the -8\% uniaxial strain
case along the $a$-direction is detailed in Sec D of the SI, as it
induces a notable shift of the direct band gap from the $\Gamma$
to the M point at the HSE06 level.}

\paragraph{\textcolor{black}{Implications of Dark Excitons and Carrier Dynamics}}

\textcolor{black}{The ''optically dark'' nature of the fundamental
band gap in the $s-C_{3}N_{6}$ monolayer provides a strategic advantage
by protecting carriers from rapid radiative recombination, thereby
extending their lifetimes. This extended lifetime is crucial for overcoming
the slow kinetics of water-splitting redox reactions, as it allows
charge carriers more time to diffuse to the surface. This is further
complemented by the strain-induced modulation of the exciton binding
energy ($E_{b}$). Our results indicate that under compressive strain,
the reduction in $m_{e}^{*}$ leads to a lower reduced mass ($\mu=\frac{m_{e}^{*}m_{h}^{*}}{m_{e}^{*}+m_{h}^{*}}$).
Additionally, the increased electronic density and lattice packing
under compression enhance the dielectric screening ($\varepsilon$)
of the Coulomb interaction. According to the Wannier-Mott model, where
$E_{b}\propto\frac{\mu}{\varepsilon^{2}}$, both the decreased $\mu$
and the increased $\varepsilon$ contribute to a significantly lower
$E_{b}$. This reduction in $E_{b}$ facilitates the thermal or interfacial
dissociation of dark excitons into free charge carriers at the material-water
interface. Consequently, the combination of a symmetry-protected dark
gap and reduced exciton binding energy under -8\% and -10\% biaxial
strain provides a synergistic pathway for enhancing the photocatalytic
quantum yield, ensuring that long-lived carriers are readily available
for the subsequent redox reactions.}

\subsubsection{\textcolor{black}{Photocatalytic mechanism and experimental design
strategies}}

\textcolor{black}{To evaluate the kinetic feasibility of the water-splitting
process for -8\% and -10\% biaxially strained $s-C_{3}N_{6}$ monolayer,
we performed a rigorous thermodynamic analysis of the HER and OER
using the Computational Hydrogen Electrode (CHE) model \cite{Man2011}.
As detailed in Sec E of the SI \cite{Yang2020,Ooka2021}, the adsorption
of intermediates ($H^{*},OH^{*},O^{*},OOH^{*}$) on the $s-C_{3}N_{6}$
surface results in excessively strong covalent bonding with the carbon
or nitrogen atoms. This leads to large negative Gibbs free energies
and prohibitive overpotentials that effectively deactivate the considered
strained $s-C_{3}N_{6}$ framework for direct catalysis. These findings
indicate that while the strained monolayer satisfies the thermodynamic
band alignment criteria for water splitting and serves as an ideal
candidate for generating long-lived charge carriers, the actual surface
redox kinetics must be facilitated by the integration of a co-catalyst.
In this context, Yang }\textit{\textcolor{black}{et al.}}\textcolor{black}{{}
\cite{Yang2013} emphasize that the primary role of a co-catalyst
is to provide specific active sites that lower the overpotential for
surface redox reactions while simultaneously promoting the spatial
separation of photogenerated carriers. Beyond co-catalyst integration,
introducing structural defects, such as nitrogen vacancies, represents
a viable alternative to tune the local electronic environment and
enhance catalytic activity, as demonstrated in our work on nitrogen-deficient
$hg-C_{3}N_{4}$ quantum dots \cite{Dange2024}.}

\textcolor{black}{By integrating $s-C_{3}N_{6}$ with appropriate
co-catalysts, such as noble metals (Pt, Ru) or earth-abundant single-atom
catalysts (Co, Ni), the charge-generation process is spatially decoupled
from the surface redox chemistry. The long-lived dark excitons generated
in the $s-C_{3}N_{6}$ provide a steady flux of carriers that can
be efficiently extracted by the co-catalyst. The co-catalyst then
provides the optimized active sites necessary to lower the overpotentials
and prevent the lattice buckling observed in the strained $s-C_{3}N_{6}$
monolayer. This synergistic strategy ensures that the superior electronic
and optical properties of the -8\% and -10\% biaxially strained $s-C_{3}N_{6}$
monolayer are fully utilized while maintaining the structural integrity
and kinetic viability of the overall photocatalytic device.}

\section{\textcolor{black}{Conclusion}}

\textcolor{black}{In summary, DFT calculations demonstrate that strain
engineering provides a realistic and effective route to activate the
photocatalytic functionality of the two-dimensional s-C\textsubscript{3}N\textsubscript{6}
monolayer. The pristine system is shown to be mechanically and thermally
stable and possesses a direct HSE06 band gap of 2.62 eV. However,
its intrinsic band-edge alignment is unfavorable for spontaneous hydrogen
evolution, preventing overall water splitting. By systematically applying
in-plane strain, we identify compressive biaxial strain in the range
of \textminus 8\% to \textminus 10\% as the only configurations that
simultaneously satisfies the thermodynamic requirements for both hydrogen
and oxygen evolution reactions. Importantly, these strained states
are found to be mechanically robust, exhibiting enhanced elastic stiffness
under compression, and thermally stable at elevated temperatures.
The magnitude of the required strain and the corresponding stresses
are within experimentally accessible limits, indicating that the proposed
photocatalytically active phase can be realized using established
techniques such as substrate-induced strain or high-pressure methods.
A significant finding of this work is the interplay between electronic
structure and carrier dynamics. Our analysis of optical properties
reveals that the fundamental band gap is optically inactive due to
symmetry-forbidden transitions, resulting in an absorption onset at
higher energies governed by deeper valence states. Under biaxial compressive
strains (-8\% and \textminus 10\%), the energy of the first allowed
optical transition exhibits a clear blue shift, despite a reduction
in the fundamental electronic band gap. This behavior reflects the
strain-induced modification of the relevant transition energies, while
the underlying selection rules remain unchanged. Although the absorption
lies primarily in the ultraviolet region, these changes occur without
compromising the thermodynamic band alignment required for photocatalytic
activity. Furthermore, the dark excitons coupled with a strain-induced
carrier mobility mismatch (arising from divergent effective mass trends),
facilitate efficient charge separation and longer radiative lifetimes.
However, our investigation into the surface kinetics highlights a
critical issue because of which the strained $s-C_{3}N_{6}$ framework
is effectively deactivated for direct catalysis due to the strong
chemisorption of reaction intermediates. The resulting large overpotentials
demonstrate that the material is chemically too vulnerable to act
as a standalone catalyst.}

\textcolor{black}{Therefore, we conclude that the optimal application
for the strained $s-C_{3}N_{6}$ monolayer is as a charge generator
in a co-catalyzed system. Decoupling the light-harvesting role from
surface redox kinetics permits the exploitation of the monolayer's
thermodynamic potential while bypassing the kinetic barriers of the
strained surface through the introduction of co-catalysts. This holistic
approach, bridging thermodynamics, exciton physics, and surface kinetics,
provides practical design guidelines for the next generation of 2D
photocatalytic devices.}

\section*{\textcolor{black}{Supporting Information}}

\textcolor{black}{The Supporting Information is available free of
charge at}

\textcolor{black}{It includes the following contents:}

\textcolor{black}{1. Electronic band structures, band alignment plots,
and results of AIMD simulations for different strained configurations
of the $s-C_{3}N_{6}$ monolayer.}

\textcolor{black}{2. Detailed optical response under uniaxial strain
along with plots of dielectric function for -8\% and -10\% biaxially
strained monolayers.}

\textcolor{black}{3. Study of surface kinetics for HER and OER}

\section*{\textcolor{black}{Author Information }}

\subsection*{\textcolor{black}{Corresponding Authors}}

\textcolor{black}{Alok Shukla:  {*}E-mail: shukla@phy.iitb.ac.in}

\subsection*{\textcolor{black}{Author Contributions}}

\textcolor{black}{The manuscript was written through contributions
of all authors. All authors have given approval to the final version
of the manuscript.}

\section*{\textcolor{black}{Notes}}

\textcolor{black}{The authors declare no competing financial interest.}
\begin{acknowledgement}
\textcolor{black}{One of the authors, K.D. acknowledges financial
assistance from the Prime Minister Research Fellowship (PMRF ID-1302054),
MHRD, India, and Space Time computational facility of Indian Institute
of Technology Bombay.}
\end{acknowledgement}

\section*{\textcolor{black}{Data availability statement}}

\textcolor{black}{Majority of the data is given in the Supporting
information and rest of the data will be provided on request.}

\textcolor{black}{\bibliography{Ref}
}
\end{document}


\begin{center}
{\Large\textbf{Supporting Information}}{\Large\par}
\par\end{center}
\title{Strain-Engineered s-C\textsubscript{3}N\textsubscript{6} Monolayer
for Efficient Water Splitting: A first-principles study}
\author{Khushboo Dange}
\address{khushboodange@gmail.com}
\affiliation{Department of Physics, Indian Institute of Technology Bombay, Powai,
Mumbai 400076, India}
\author{Alok Shukla{*}}
\address{shukla@iitb.ac.in}
\affiliation{Department of Physics, Indian Institute of Technology Bombay, Powai,
Mumbai 400076, India}
\maketitle

\subsection{Electronic band structures of the $s-C_{3}N_{6}$ monolayer under
application of strains}

\begin{figure}[H]
\begin{centering}
\includegraphics[scale=0.45]{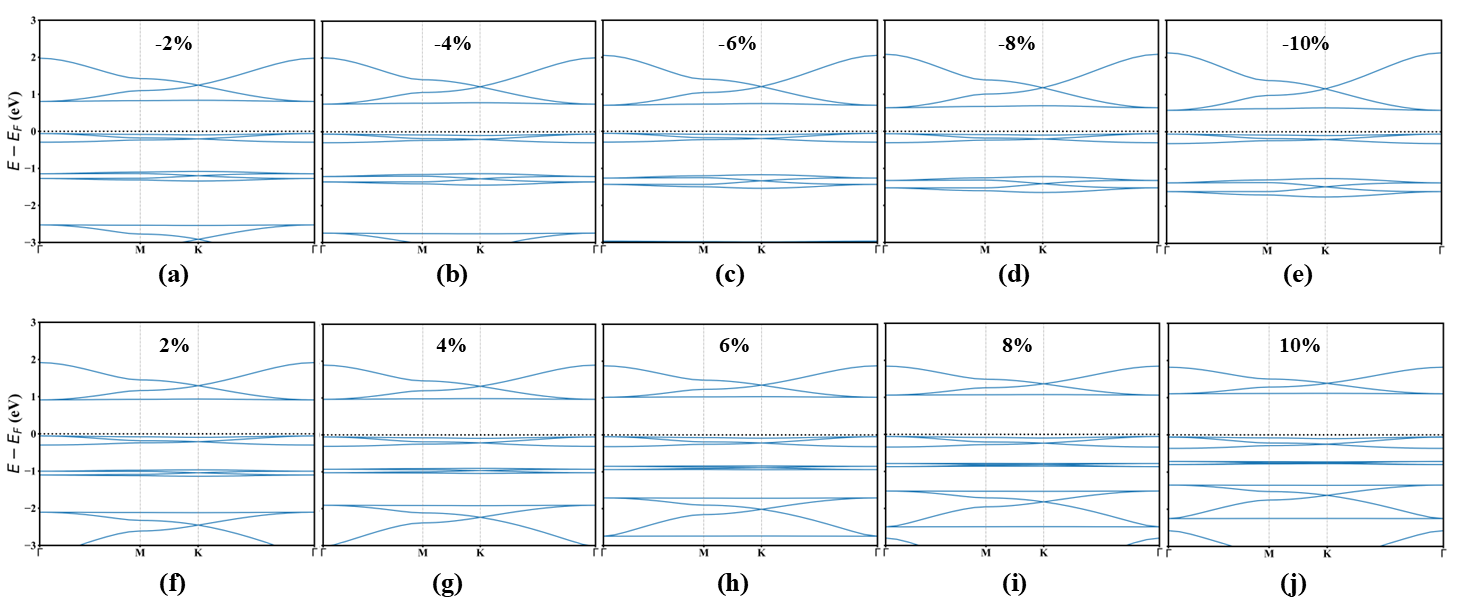}
\par\end{centering}
\caption{Calculated GGA-PBE band structures for the s-C\protect\textsubscript{3}N\protect\textsubscript{6}
monolayer under biaxial strain.}

\end{figure}

\begin{figure}[H]
\begin{centering}
\includegraphics[scale=0.45]{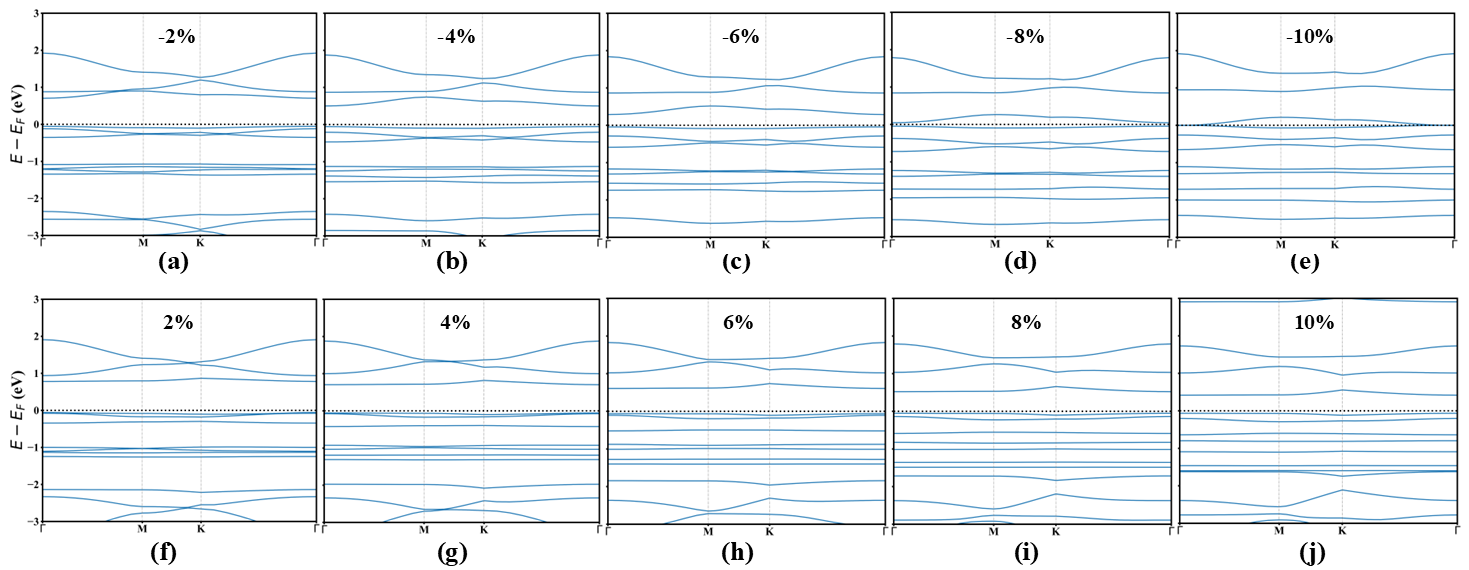}
\par\end{centering}
\caption{Calculated GGA-PBE band structures for the s-C\protect\textsubscript{3}N\protect\textsubscript{6}
monolayer under the effect of uniaxial strains along $a$-direction.}
\end{figure}

\begin{figure}[H]
\begin{centering}
\includegraphics[scale=0.45]{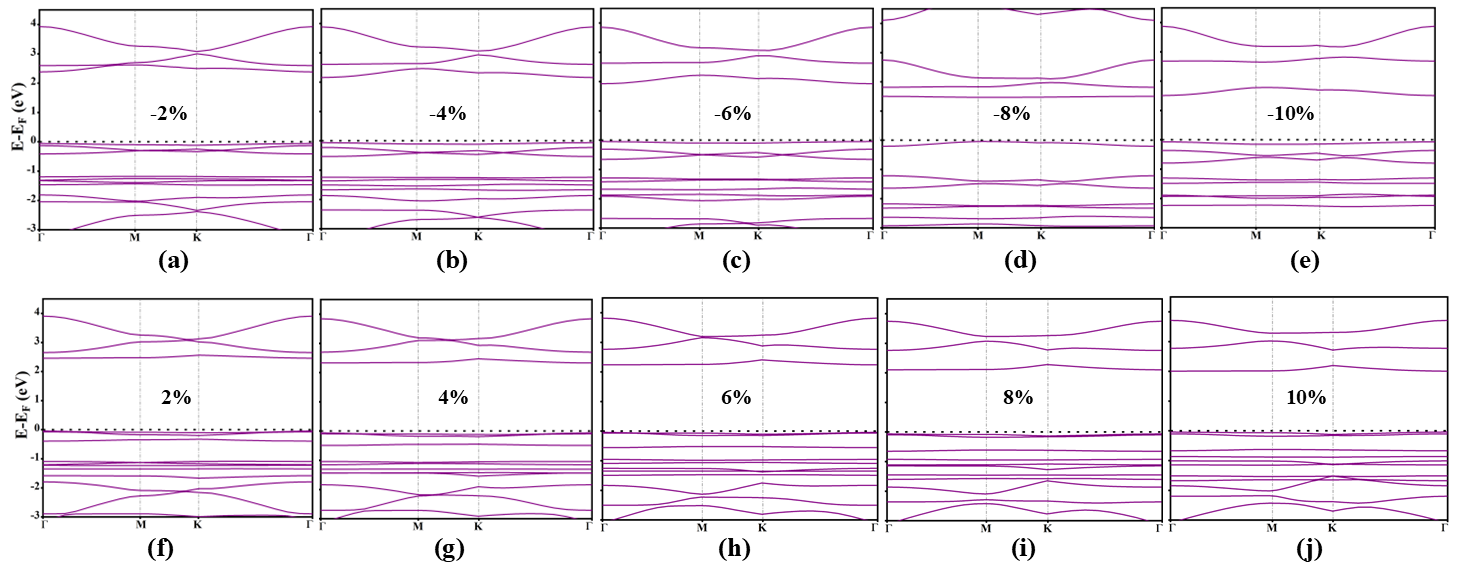}
\par\end{centering}
\caption{Calculated HSE06 band structures for the s-C\protect\textsubscript{3}N\protect\textsubscript{6}
monolayer under uniaxial strains along the $a$-direction.}
\end{figure}
\begin{figure}[H]
\begin{centering}
\includegraphics[scale=0.45]{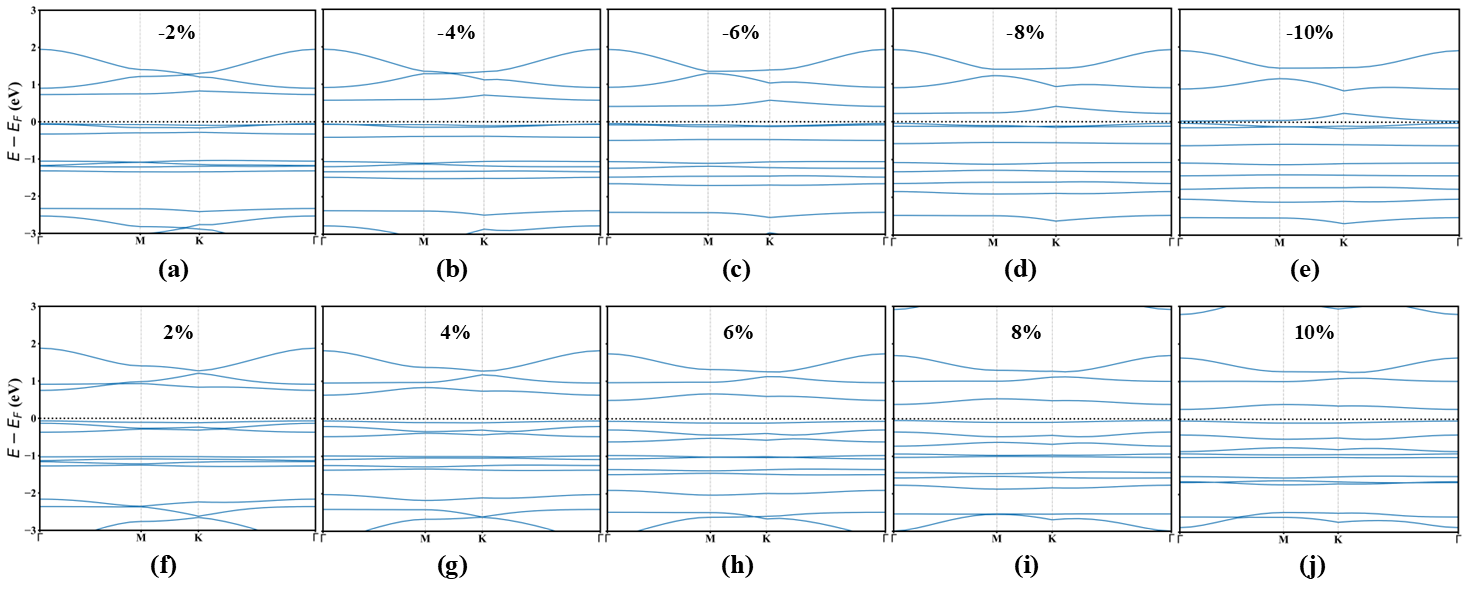}
\par\end{centering}
\caption{Calculated GGA-PBE band structures for the s-C\protect\textsubscript{3}N\protect\textsubscript{6}
monolayer under uniaxial strains along the $b$-direction.}
\end{figure}

\begin{figure}[H]
\begin{centering}
\includegraphics[scale=0.45]{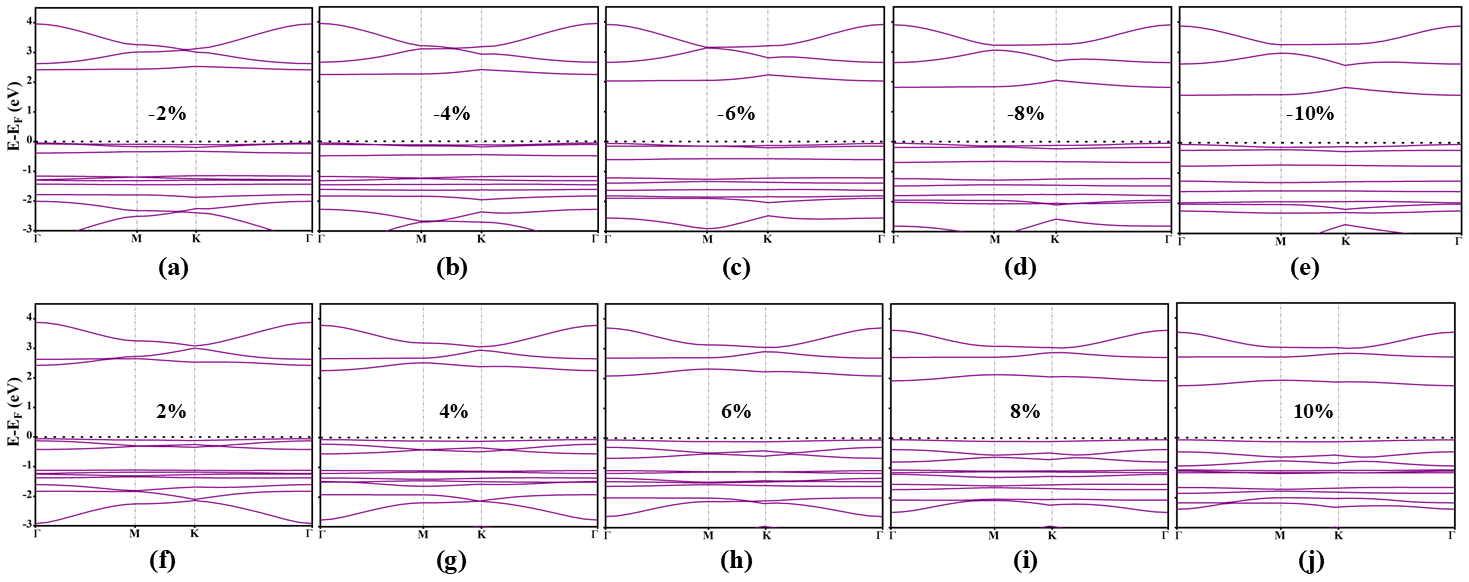}
\par\end{centering}
\caption{Calculated HSE06 band structures for the s-C\protect\textsubscript{3}N\protect\textsubscript{6}
monolayer under the effect of uniaxial strains along $b$-direction.}

\end{figure}

\subsection{Band alignment for uniaxially strained $s-C_{3}N_{6}$ monolayer}

\begin{figure}[H]
\begin{centering}
\includegraphics[scale=0.5]{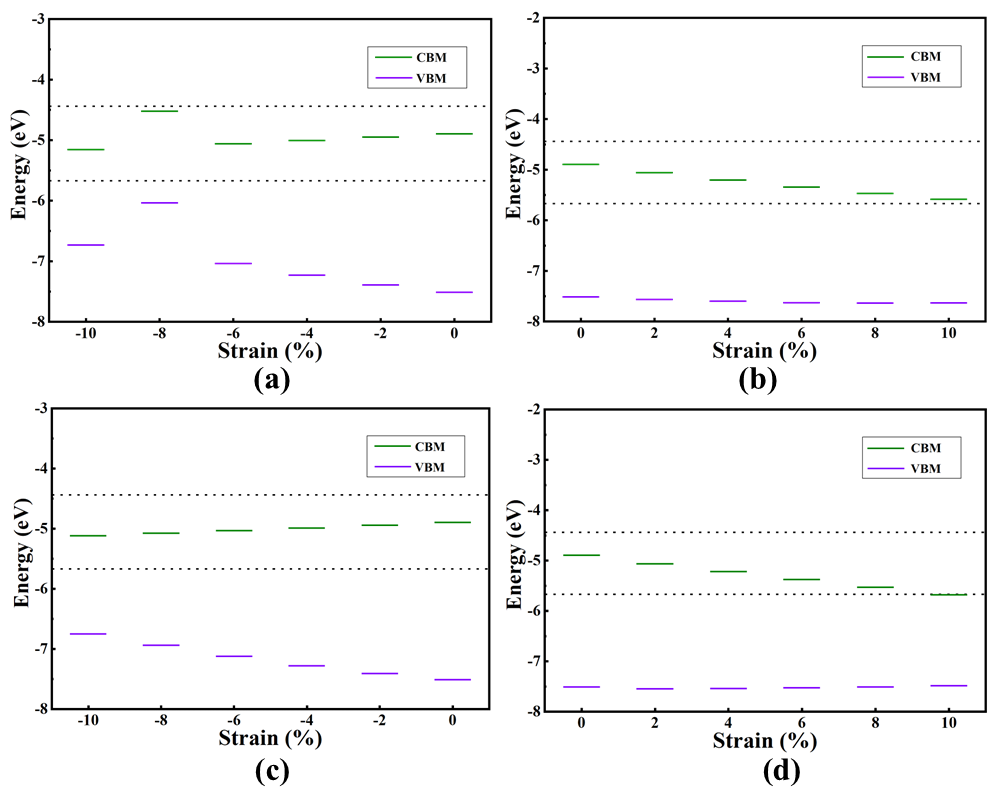}
\par\end{centering}
\caption{CBM and VBM positions of the s-C\protect\textsubscript{3}N\protect\textsubscript{6}
monolayer under the effect of uniaxial strains: (a) compressive and
(b) tensile along the $a$-direction, (c) compressive and (d) tensile
along the $b$-direction.}
\end{figure}

\subsection{Ab initio molecular dynamics (AIMD) simulations for -8\% and -10\%
biaxial strain}

\begin{figure}[H]
\begin{centering}
\includegraphics[scale=0.5]{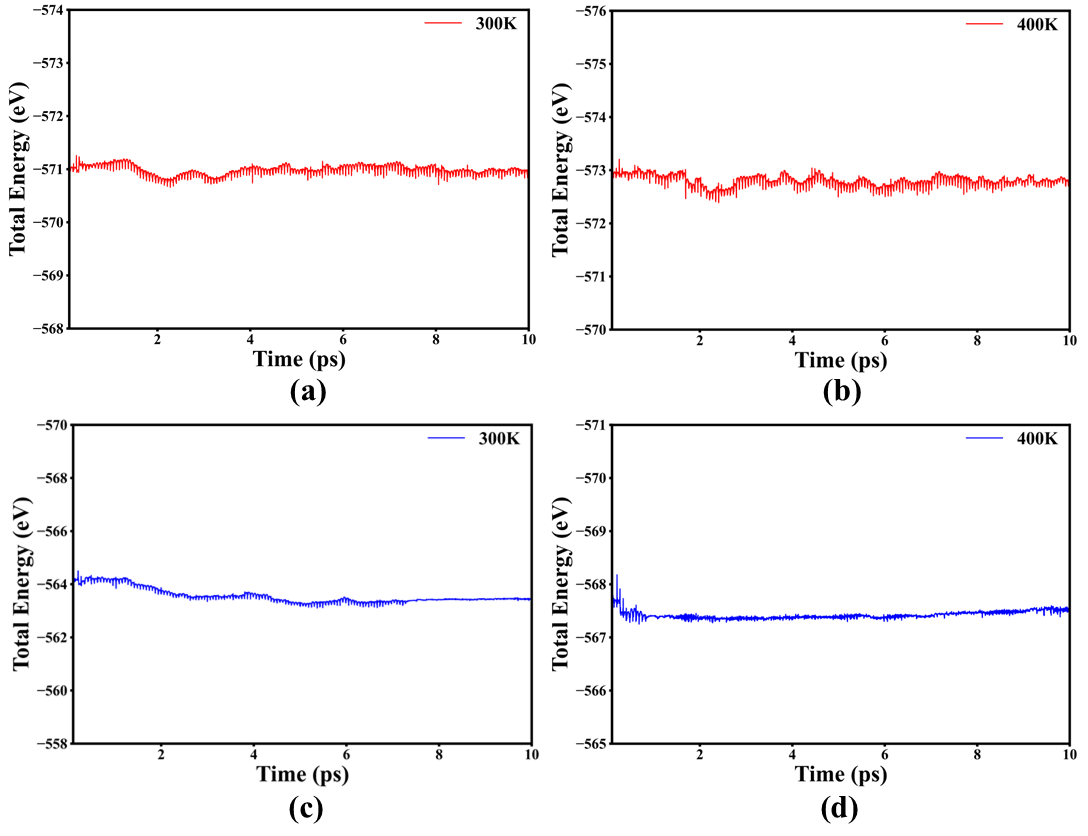}
\par\end{centering}
\caption{The variation of the total energy of the biaxially strained s-C\protect\textsubscript{3}N\protect\textsubscript{6}
monolayer as a function of the simulation time for: (a) \& (b) -8
\% and (c) \&(d) -10 \% biaxial strains, at 300 K \& 400 K, respectively. }

\end{figure}

\subsection{Optical properties}

\begin{figure}[H]
\begin{centering}
\includegraphics[scale=0.5]{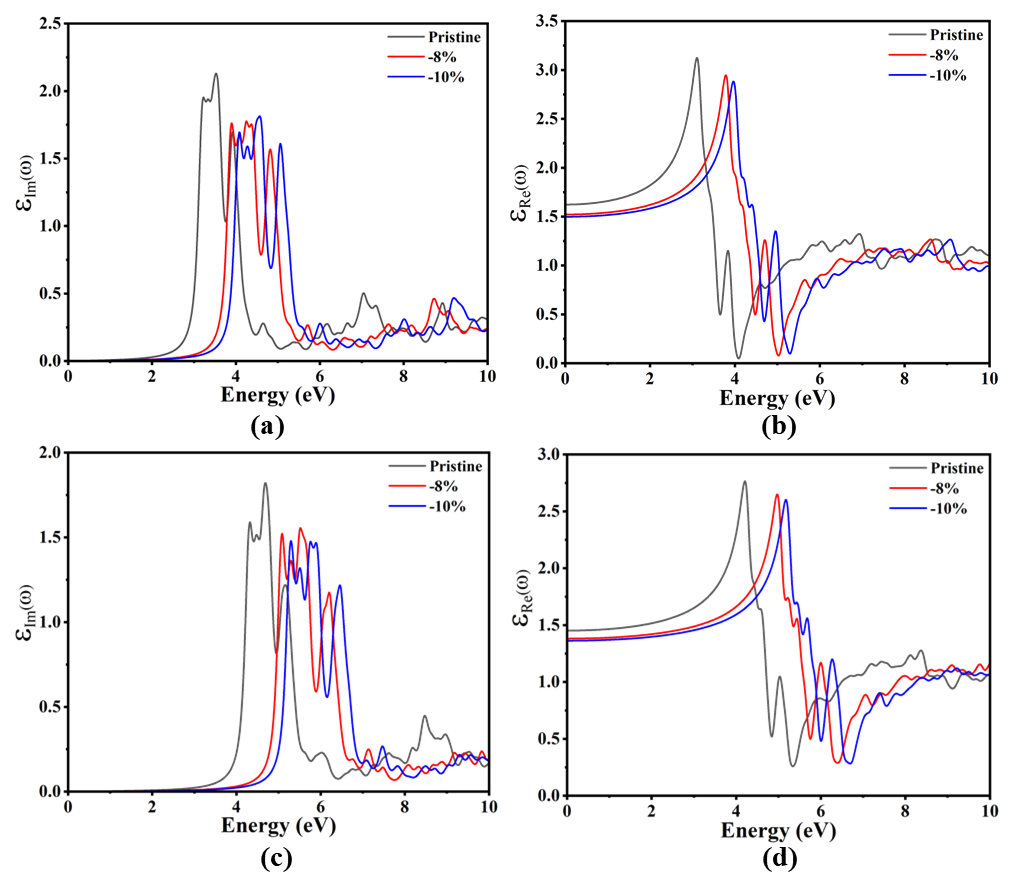}
\par\end{centering}
\caption{(a) \& (c) Imaginary and (b) \& (d) real parts of the dielectric function
as a function of energy calculated using the GGA-PBE \& HSE06 functionals,
respectively.}

\end{figure}

\subsubsection*{Optical response under uniaxial strain}

The optical properties under uniaxial strain are examined for the
representative case of \textminus 8\% strain applied along the $a$-direction.
This configuration is particularly noteworthy because, unlike all
other uniaxial strain cases where the direct band gap remains at the
$\Gamma$ point, it induces a shift of the direct band gap to the
M point at the HSE06 level. This momentum-space relocation provides
an opportunity to test the robustness of the optical selection rules. 

\begin{figure}
\begin{centering}
\includegraphics[scale=0.4]{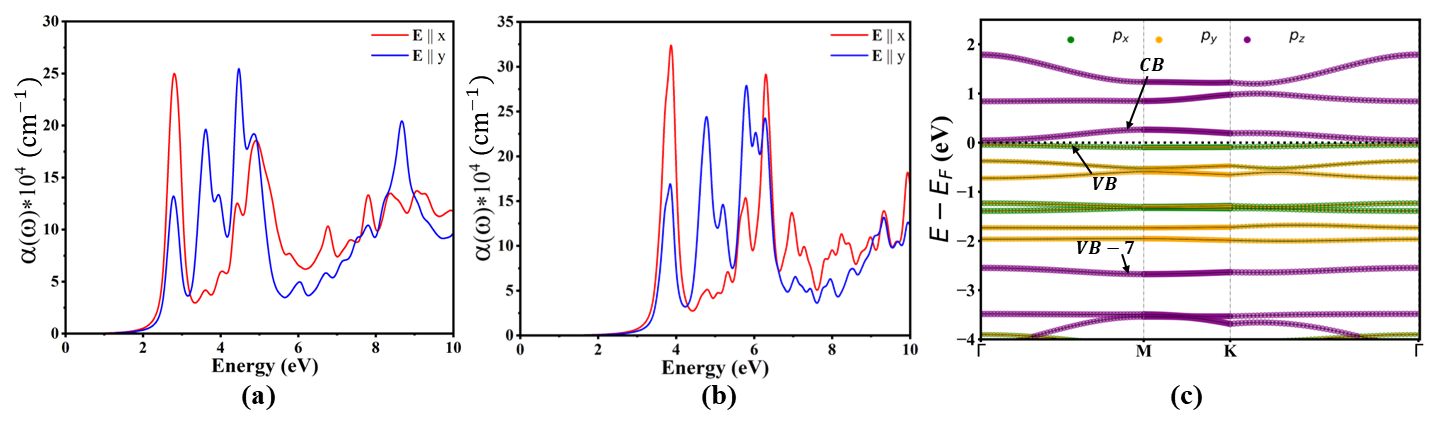}
\par\end{centering}
\caption{\protect\label{fig:Optical-absorption}Polarized optical absorption
coefficient $\alpha(\omega)$ of the s-C\protect\textsubscript{3}N\protect\textsubscript{6}
monolayer under the effect of -8\% uniaxial strain along $a$-direction,
calculated using (a) GGA-PBE and (b) HSE06 functionals. (c) $lm$-decomposed
GGA-PBE band structure showing the contribution of the $p$-orbitals
of nitrogen atoms.}

\end{figure}

The optical absorption coefficient $\alpha(\omega)$, computed using
both the GGA-PBE and HSE06 functionals, is shown in Fig. \ref{fig:Optical-absorption}(a)
and (b), respectively. The application of uniaxial strain lowers the
crystal symmetry from hexagonal $P6/m$ to monoclinic $P2/m$, resulting
in anisotropic optical behavior. Consequently, the absorption spectra
for light polarized along the $x-$direction (E||$x$) and $y-$direction
(E||$y$) become distinct, reflecting the broken in-plane symmetry.
Despite the relocation of the direct band gap and the substantial
reduction of the electronic gap ($E_{g}$), which decreases to as
low as 0.09 eV at the PBE level, the optical absorption onset remains
significantly higher in energy. The first pronounced absorption peak
appears at 2.79 eV at the PBE level and is identical for both polarization
directions. A similar trend is observed at the HSE06 level, where
the first absorption peak occurs at 3.85 eV, far above the $E_{g}$
of 1.5 eV. This indicates that the fundamental band-edge transition
remains optically inactive even after the shift of the direct gap
from $\Gamma$ to M.

To understand this behavior, the orbital-resolved PBE band structure
for the \textminus 8\% uniaxial strain case is analyzed, as shown
in Fig. \ref{fig:Optical-absorption}(c). The topmost valence band
(VB) is found to be dominated by the in-plane $p_{x}$ orbitals, whereas
the lowest conduction band (CB) retains its out-of-plane $p_{z}$
character. As a result, the $VB\rightarrow CB$ transition which corresponds
to a $\sigma\rightarrow\pi^{*}$ excitation remains forbidden. This
explains why the fundamental gap remains optically dark. Consistent
with the pristine and biaxially strained cases, the first optically
allowed transition originates from deeper valence states. In particular,
the VB-7 band, which possesses $p_{z}$ character, gives rise to the
first dipole-allowed transition to the CB with maximum intensity at
the $\Gamma$ point due to the minimum energy separation between the
CB and VB\textminus 7 states. Since the HSE06 functional preserves
the orbital character and primarily induces a rigid shift of the energy
levels, this $\pi\rightarrow\pi^{*}$ transition consistently defines
the first optically allowed excitation.

These results demonstrate that the optical selection rules are primarily
dictated by the orbital character of the band-edge states rather than
their position in momentum space. Consequently, the shift of the direct
band gap from $\Gamma$ to M at the HSE06 level does not alter the
optical activity of the fundamental transition. The system thus remains
optically inactive at the band edge, with the absorption occurring
only at higher energies, where the symmetry-allowed transitions are
present.

\subsection{Thermodynamic framework and catalytic pathways}

The overall photocatalytic water splitting process is governed by
the two coupled half-reactions: hydrogen evolution reaction (HER)
and oxygen evolution reaction (OER). To assess the catalytic potential
of the -8\% and -10\% biaxially strained $s-C_{3}N_{6}$ monolayers,
we employed the Computational Hydrogen Electrode (CHE) model \citep{Manetal2011}.
Within this framework, the catalytic efficiency is determined by the
change in Gibbs free energy ($\Delta G$) for each elementary reaction
step. This value is calculated as 

\begin{equation}
\Delta G=\Delta E^{ads}+\Delta E_{ZPE}-T\Delta S\label{eq:general-Gibb's}
\end{equation}

where the term $\Delta E^{ads}$ accounts for the adsorption energy
of the specific intermediates ($H^{*},OH^{*},O^{*},OOH^{*}$), while
the remaining terms represent zero-point energy ($\Delta E_{ZPE}$)
and entropy corrections ($T\Delta S$). The HER is a two-electron
reduction process that takes place at the conduction band edge. It
involves two elementary steps: the initial adsorption of a proton
to form an $H^{*}$ intermediate, followed by the release of $H_{2}$,
i.e.,
\begin{equation}
H^{+}+e^{-}+*\rightarrow H^{*}
\end{equation}

\begin{equation}
H^{*}+H^{+}+e^{-}\rightarrow H_{2}(g)+*
\end{equation}

where $*$ denotes an active catalytic site. By applying the entropic
and zero-point energy corrections used in various studies \citep{Dange2024},
Eq. \ref{eq:Gibb's=000020HER} for HER becomes: 
\begin{equation}
\Delta G_{H^{*}}=\Delta E_{H^{*}}^{ads}+0.24\label{eq:Gibb's=000020HER}
\end{equation}

The resulting overpotential ($\eta_{HER}$), calculated as 
\begin{equation}
\eta_{HER}=-\frac{\left|\Delta G_{H^{*}}\right|}{e}\label{eq:overpot-HER}
\end{equation}
serves as a direct indicator of catalytic activity, with a negative
sign accounting for the cathodic nature of the reaction. 

\begin{figure}[H]
\begin{centering}
\includegraphics[scale=0.5]{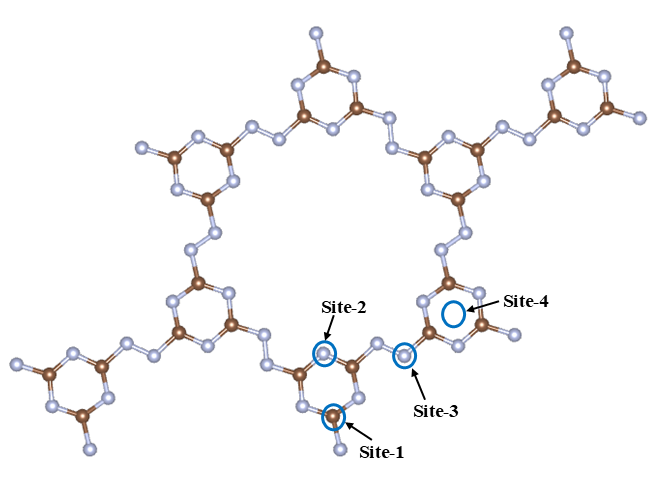}
\par\end{centering}
\caption{\protect\label{fig:Four-different-sites}Four different sites for
adsorption (labeled as site-1 to site-4) of HER and OER reaction intermediates
($H^{*},OH^{*},O^{*},OOH^{*}$) encircled in blue.}
\end{figure}

To assess the site-dependent catalytic performance, we considered
four distinct adsorption sites (see Fig. \ref{fig:Four-different-sites}),
each providing a unique chemical environment: the top of the C atom
(site-1), N1 atom (site-2), N2 atom (site-3), and the hole of the
triazine ring (site-4). Our calculations across these sites reveal
that $H^{*}$ forms an excessively strong covalent bond with the surface
atoms, leading to significant structural distortion and buckling of
the $s-C_{3}N_{6}$ framework. This intense interaction is reflected
in the large negative adsorption energies ($\Delta E_{H^{*}}^{ads}$),
which translate to $\Delta G_{H^{*}}(\eta_{HER})$ values ranging
from -3.06 to -4.48 eV (-3.06 to -4.48) for the -8\% strained case
and -3.95 to -6.84 eV (-3.95 to -6.84) for the -10\% strained case
(exact values listed in Table \ref{tab:Calculated-adsorption-energiesHER}).
According to the Sabatier principle \citep{Ooka2021}, an ideal catalyst
should bind intermediates neither too strongly nor too weakly. The
extreme negative free energies and overpotentials observed here indicate
that the $s-C_{3}N_{6}$ surface becomes inhibited by the $H^{*}$
intermediate, as the energy required for the subsequent desorption
of $H_{2}$ is prohibitively high. These findings confirm that while
the biaxially strained $s-C_{3}N_{6}$ monolayer satisfies the thermodynamic
requirements for water splitting, its intrinsic surface kinetics for
HER are highly unfavorable, necessitating the integration of a co-catalyst
to facilitate the reaction.

\begin{table}[H]
\caption{\protect\label{tab:Calculated-adsorption-energiesHER}Calculated adsorption
energies ($\Delta E_{ads}$), change in Gibb's free energies ($\Delta G_{H^{*}}$)
in eV, and overpotential ($\eta_{HER}$) for HER corresponding to
four different adsorption sites of -8\% and -10\% biaxially strained
$s-C_{3}N_{6}$ monolayer.}

\centering{}%
\begin{tabular}{ccccccc}
\toprule 
\multirow{2}{*}{Sites} & \multicolumn{3}{c}{-8\%} & \multicolumn{3}{c}{-10\%}\tabularnewline
\cmidrule{2-7}
 & $\Delta E_{H^{*}}^{ads}$ & $\Delta G_{H^{*}}$ & $\eta_{HER}$ & $\Delta E_{H^{*}}^{ads}$(eV) & $\Delta G_{H^{*}}$ & $\eta_{HER}$\tabularnewline
\midrule
\midrule 
1 & -3.30 & -3.06 & -3.06 & -4.19 & -3.95 & -3.95\tabularnewline
\midrule 
2 & -4.43 & -4.19 & -4.19 & -7.08 & -6.84 & -6.84\tabularnewline
\midrule 
3 & -4.72 & -4.48 & -4.48 & -7.05 & -6.82 & -6.82\tabularnewline
\midrule 
4 & -4.43 & -4.19 & -4.19 & -6.69 & -6.45 & -6.45\tabularnewline
\bottomrule
\end{tabular}
\end{table}

In contrast to the HER, OER is a more complex four-electron oxidation
process occurring at the valence band edge. It involves the sequential
formation of oxygenated intermediates through the following elementary
steps:

\begin{equation}
*+H_{2}O(l)\rightarrow OH^{*}+(H^{+}+e^{-})
\end{equation}

\begin{equation}
OH^{*}\rightarrow O^{*}+(H^{+}+e^{-})
\end{equation}

\begin{equation}
O^{*}+H_{2}O(l)\rightarrow OOH^{*}+(H^{+}+e^{-})
\end{equation}

\begin{equation}
OOH^{*}\rightarrow*+O_{2}(g)+(H^{+}+e^{-})
\end{equation}

The Gibbs free energies for these intermediates are obtained using
standard energy corrections \citep{Yang2020} :
\begin{equation}
\Delta G_{OH^{*}}=\Delta E_{OH^{*}}^{ads}+0.35
\end{equation}

\begin{equation}
\Delta G_{H^{*}}=\Delta E_{O^{*}}^{ads}+0.05
\end{equation}

\begin{equation}
\Delta G_{OOH^{*}}=\Delta E_{OOH^{*}}^{ads}+0.40
\end{equation}

The step wise free energies ($\Delta G_{1}$ to $\Delta G_{3}$) were
then derived from these values as: 
\begin{equation}
\Delta G_{1}=\Delta G_{OH^{*}}\label{eq:G1-oer}
\end{equation}

\begin{equation}
\Delta G_{2}=\Delta G_{O^{*}}-\Delta G_{OH^{*}}\label{eq:G2-oer}
\end{equation}

\begin{equation}
\Delta G_{3}=\Delta G_{OOH^{*}}-\Delta G_{O^{*}}\label{eq:G3-oer}
\end{equation}

To ensure thermodynamic consistency and avoid the well-known errors
associated with the direct calculation of the $O_{2}$ gas-phase energy,
the final step is evaluated as
\begin{equation}
\Delta G_{4}=4.92-(\Delta G_{1}+\Delta G_{2}+\Delta G_{3})\label{eq:G4-oer}
\end{equation}

where 4.92 eV represents the total free energy change for water splitting.
The overpotential ($\eta_{OER}$) for this half-reaction is determined
by the potential-determining step (PDS), the elementary transition
with the largest $\Delta G$ calculated as 
\begin{equation}
\eta_{OER}=\frac{max(\Delta G_{1},\Delta G_{2},\Delta G_{3},\Delta G_{4})}{e}-1.23\label{eq:overp-oer}
\end{equation}

For both the -8\% and -10\% biaxially strained monolayers, our calculations
across the four symmetry-unique adsorption sites (site-1 to site-4)
reveal extreme deviations from the ideal thermodynamic requirement.
As shown in Table \ref{tab:Change-in-Gibb'sOER}, the step-wise free
energies in most cases exhibit large positive or negative values,
indicating that the binding of oxygenated intermediates is either
excessively strong or prohibitively weak. These results lead to large
overpotentials that severely restrict OER performance, placing the
pristine surface far from the top of the OER volcano plot. Beyond
these thermodynamic barriers, the mechanical vulnerability of the
$s-C_{3}N_{6}$ framework under OER conditions is a major concern.
The intense chemisorption of oxygenated species exerts such significant
mechanical stress that the lattice is observed to buckle or even rupture
during optimization. This structural collapse proves that the considered
biaxially strained $s-C_{3}N_{6}$ frameworks is chemically too vulnerable
to serve as the primary site for these reactions, as the intermediates
physically destroy the framework they are supposed to react upon.
It is important to note that certain configurations show a degree
of mechanical stability. Specifically, for the -8\% biaxially strained
surface, adsorption at site-2 and site-3 results only in slight lattice
buckling rather than complete rupture. In these instances, the binding
of the oxygenated intermediates is comparatively weaker than in other
configurations, which leads to a reduced $\eta_{OER}$ of 1.80 eV.
However, within the context of electrochemistry, a 1.80 eV overpotential
remains prohibitively large for OER to proceed with any practical
efficiency. 

Furthermore, the feasibility of total water splitting depends on the
synchronized performance of both half-reactions, as the HER and OER
must occur simultaneously and at an equivalent rate to maintain charge
neutrality. Even if we consider the most favorable overpotential values,
the systemic imbalance is stark. The minimum overpotential obtained
for HER across all considered cases is significantly large, with $\eta_{HER}$
reaching -3.06 eV. When paired with the reduced $\eta_{OER}$ of 1.80
eV, the total energy barrier for the combined process far exceeds
the available driving force provided by the band-edge positions. Consequently,
the biaxially strained $s-C_{3}N_{6}$ monolayer is a highly efficient
candidate for water splitting because it strictly satisfies the thermodynamic
band alignment requirements. However, while the material excels at
providing the necessary potential for spontaneous reactions, the actual
surface redox kinetics present a significant bottleneck due to the
excessive binding of intermediates. Therefore, to facilitate the HER
and OER with minimum required overpotential and ensuring structural
stability, the integration of suitable co-catalysts is required to
bridge the gap between thermodynamic potential and kinetic feasibility.

\begin{table}[H]
\caption{\protect\label{tab:Change-in-Gibb'sOER}Change in Gibb's free energies
($\Delta G_{1},\Delta G_{2},\Delta G_{3},\Delta G_{4}$) in eV for
intermediate OER steps and corresponding overpotential ($\eta$) for
considered four different adsorption sites of -8\% and -10\% biaxially
strained $s-C_{3}N_{6}$ monolayer. }

\centering{}%
\begin{tabular}{ccccccccccc}
\toprule 
\multirow{2}{*}{Sites} & \multicolumn{5}{c}{-8\%} & \multicolumn{5}{c}{-10\%}\tabularnewline
\cmidrule{2-11}
 & $\Delta G_{1}$ & $\Delta G_{2}$ & $\Delta G_{3}$ & $\Delta G_{4}$ & $\eta_{OER}$ & $\Delta G_{1}$ & $\Delta G_{2}$ & $\Delta G_{3}$ & $\Delta G_{4}$ & $\eta_{OER}$\tabularnewline
\midrule
\midrule 
1 & -3.66 & 1.04 & 4.15 & 3.39 & 2.92 & -6.22 & 4.39 & 0.94 & 5.81 & 4.58\tabularnewline
\midrule 
2 & -1.19 & 1.46 & 1.61 & 3.03 & 1.80 & -4.37 & 1.70 & 2.28 & 5.31 & 4.08\tabularnewline
\midrule 
3 & -1.96 & 1.89 & 1.95 & 3.03 & 1.80 & -4.33 & 1.15 & 2.76 & 5.17 & 3.94\tabularnewline
\midrule 
4 & -1.91 & 4.77 & -1.33 & 3.39 & 3.54 & -4.53 & 1.57 & 2.12 & 5.76 & 4.53\tabularnewline
\bottomrule
\end{tabular}
\end{table}

\bibliographystyle{apsrev4-2}
\bibliography{Ref_SI}